# PGMG: A Pharmacophore-Guided Deep Learning Approach for Bioactive Molecular Generation

Huimin Zhu[1,†], Renyi Zhou[1,†], Jing Tang[2] and Min Li[1,*]

[1] School of Computer Science and Engineering, Central South University, Changsha 410083, China
[2] Faculty of Medicine, University of Helsinki, Helsinki, 00290, Finland

[†] These two authors contribute equally to the work.
[*] Corresponding author, limin@mail.csu.edu.cn

**Abstract**

The rational design of novel molecules with desired bioactivity is a critical but challenging task in drug discovery, especially when treating a novel target family or understudied targets. Here, we propose PGMG, a pharmacophore-guided deep learning approach for bioactivate molecule generation. Through the guidance of pharmacophore, PGMG provides a flexible strategy to generate bioactive molecules with structural diversity in various scenarios using a trained variational autoencoder. We show that PGMG can generate molecules matching given pharmacophore models while maintaining a high level of validity, uniqueness, and novelty. In the case studies, we demonstrate the application of PGMG to generate bioactive molecules in ligand-based and structure-based drug de novo design, as well as in lead optimization scenarios. Overall, the flexibility and effectiveness of PGMG make it a useful tool for accelerating the drug discovery process.

**Introduction**

The acquisition of biologically active compounds is a vital but challenging step in drug discovery. It has been estimated that the drug-like chemical space is as large as $10^{60}$ obeying Lipinski's "Rule of Five"[1, 2]. Hence, it is an extremely difficult task to search for desired molecules in such a huge space. Traditionally, hit compounds which exhibit initial activity on a specific target can be obtained from natural products, designed by medicinal chemists, or acquired by high-throughput screening (HTS)[3]. These methods consume a lot of human and financial resources, which makes the acquisition of hit compounds inefficient and costly. Recently, some deep generative models have been proposed for the rational design of novel molecules with desired properties, which provide a new perspective for this task.

There are two primary types of molecule generation models: SMILES-based models and graph-based models. In SMILES-based models, a molecule is represented as a SMILES string. Sequence generation methods are used to generate molecules in SMILSE-based models, such as BIMODAL[4], MCMG[6], VAE[4], SMILES LSTM[5]. In graph-based models, a molecule is represented as an undirected graph $G(V, E)$, with $V$ designating heavy atoms or substructures and $E$ denoting covalent bonds, such as MGM[6], JTVAE[7], Mol-CycleGAN[8] and MolDQN[9]. Regardless of the representation of molecules, most above methods aim at generating molecules with given physicochemical properties, such as the Wildman-Crippen partition coefficient (LogP), synthetic accessibility (SA), molecular weight (molWt), quantitative estimate of drug likeness (QED), and others. However, these models are less suitable for generating bioactive molecules, which is a decisive step for drug discovery. For a specified target, these models require a large dataset of known active molecules to fine-tune and thus cannot be applied to a novel target or targets with few active compounds.

Designing molecules using deep generative models with biological activity remains challenging[10, 11, 12]. As mentioned above, one of the main obstacles is the limited data on target-specific molecules, which makes it difficult for models to learn the structure-activity relationship. For a novel target family, the paucity of activity data is even more severe. Besides, the scarcity of activity data affects the strategy of drug design. For example, the choice of ligand-based drug design or structure-based drug design depends on what information can be used, which narrows down the application scenarios of many methods. It is clear that incorporating expert knowledge in the generation process is beneficial to the full utilization of bioactivity data information[13]. Therefore, combining deep generative models with knowledge in biochemistry to efficiently use the scarce data to design biologically active molecules is a crucial project.

Up to now, some methods that generate bioactive molecules by combining prior knowledge from biochemistry into molecule generation models have been proposed. For example, conditioned GAN can be used to design active-like molecules for desired gene expression signatures[14], which provides a new perspective for molecule generation. However, the structure-activity relationship between the biological activity and the molecules generated by such methods is ambiguous. DeLinker[15] and SyntaLinker[16] adopt fragment-based drug design and retain active fragments while updating linkers to generate active molecules, and DEVELOP[13] combines DeLinker with chemical features as constraints to improve the quality of the generated molecules. The fragment-based approaches require explicit knowledge of the active fragments, which lead to a limited chemical space for the model. DeepLigBuilder[17] and 3D-Generative-SBDD[18] utilize the structure-based drug design strategy and generate molecules based on the binding sites between molecules and proteins in the 3D Euclidean Space. However, these methods are limited when the binding site or the target structure is unknown. There are also some methods that use electronic features in molecule generation. For example, Reduced Graph[19] simplifies a SMILES to an acyclic graph of functional group as its input to generation. Shape-based method proposed by Skalic et al[20] generate molecules from a 3D representation using a seed ligand with a conditional chemical features. These methods require seed compounds to collect the input electronic features. The above generative models may perform well on specific types of activity data, but their usages are limited because of their assumptions on the data types.

Here, we propose PGMG, a pharmacophore-guided molecule generation approach based on deep learning. PGMG uses pharmacophore models as a bridge to connect different types of activity data and can generate molecules with biological activity and structural diversity even for new targets or targets with few activity data. A pharmacophore is a set of chemical features that are necessary for a drug to bind to a target and can be constructed by superimposing a small number of active compounds[21] or observing the structure of a given target[22]. Traditional drug design based on pharmacophores has many successful applications[23, 24], but its potential in deep generative models has not been exploited. There are some works that use pharmacophore-like information in molecule generation, like the aforementioned Reduced Graph[19] and the shape-based design[19]. However, the pharmacophore-like features used in these methods are incomplete and can only be extracted from seed compounds, making it difficult for domain knowledge to be leveraged. A pharmacophore model consists of two parts: the pharmacophore features (such as hydrogen-bonded donors, hydrogen-bonded acceptors, aromatic rings, hydrophobic centers) and the spatial information of pharmacophore. In PGMG, we use a complete graph to fully represent a pharmacophore with each node corresponding to a pharmacophore feature and the spatial information encoded as the distance between each node pair. Given the graph as the sole input, PGMG can generate molecules matching the corresponding pharmacophore. This gives PGMG the capability to utilize different types of activity data in a uniform representation and a biologically meaningful way to control the process of the bioactivity molecule design.

Since pharmacophores and molecules have a many-to-many relationship, PGMG introduces latent variables to model such a relationship and boost the variety of generated molecules. Besides, the transformer structure is employed as the backbone to learn implicit rules of SMILES strings to map between latent variables and molecules. We evaluate the PGMG performance comprehensively in molecule generation with goal-directed metrics and drug-like metrics. The results show that PGMG can generate molecules satisfying given pharmacophore models and pharmacokinetic requirements, while maintaining a high level of validity, uniqueness, and novelty. The case studies further demonstrate that PGMG provides an effective strategy for both ligand-based and structure-based drug de novo designs and lead optimization.

## Results

### Overview of PGMG

Our proposed PGMG is a pharmacophore-guided molecular generation approach based on deep learning. The overall architecture of PGMG is illustrated in **Figure 1**.

Given a target pharmacophore, the goal of PGMG is to generate molecules which matches the pharmacophore. Here, we introduce a set of latent variables $z$ to deal with the many-to-many mapping between pharmacophores and molecules. Thus, a molecule $x$ can be represented as a unique combination of two complementary encodings including $c$ representing the given pharmacophore and $z$ corresponding to how chemical groups are placed within the molecule. From another perspective, the latent variables $z$ grant PGMG to model multiple modes in the conditional distribution

$$P(x|c) = \int_{z \sim P(Z|C)} P(x|c,z)P(z|c)dz \quad (1)$$

We train two neural networks, an encoder network $P_\phi(z|c,x)$ to approximate $P(z|c)$ indirectly and a decoder network $P_\theta(x|c,z)$ to approximate $P(x|c,z)$. We embed molecules in SMILES format into dense feature vectors and use Gated GCN[25] to embed pharmacophore models. The transformer structure proposed by Vaswani et al.[26] is used as the backbone of our model to learn the mapping between pharmacophore and molecular structures.

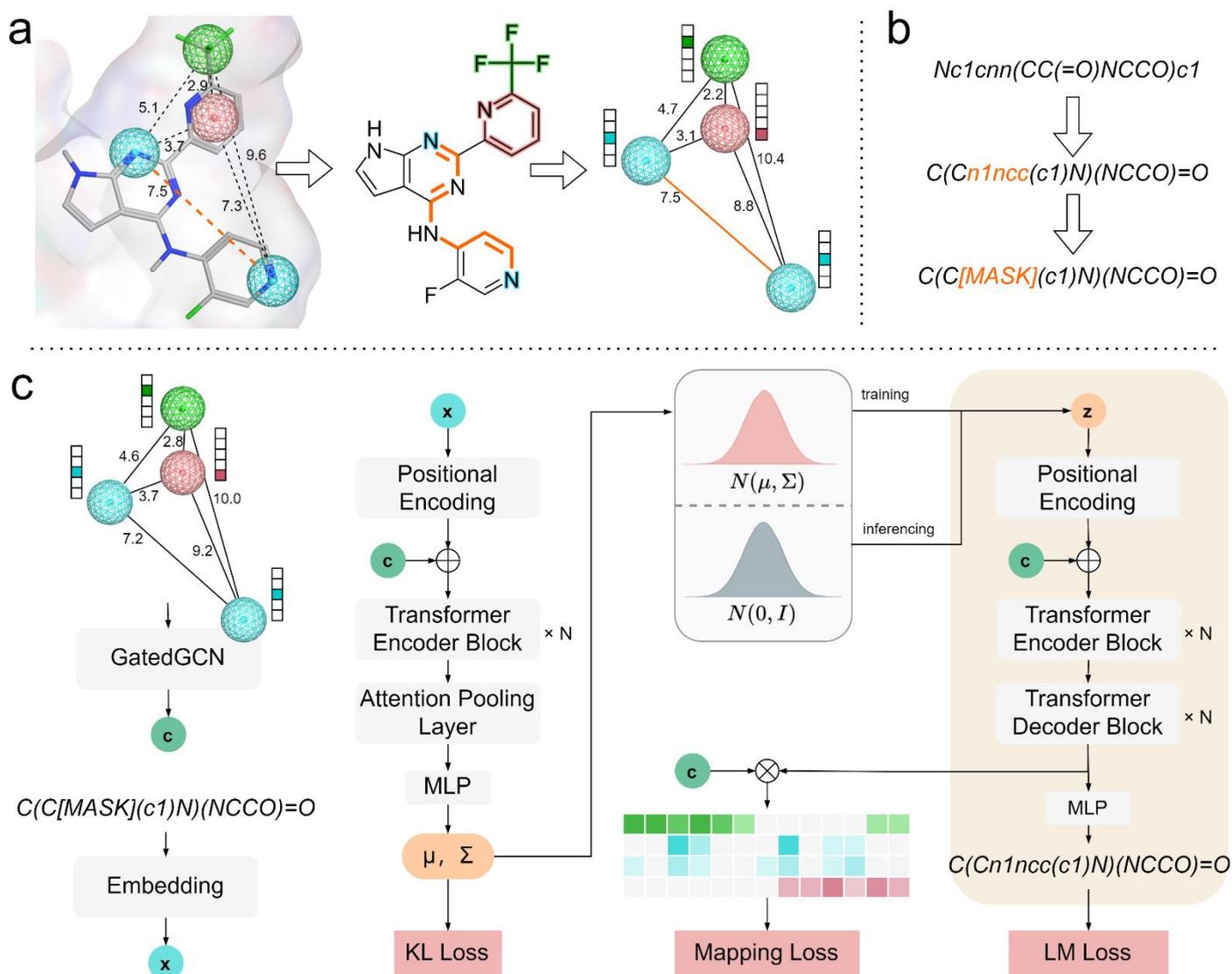

**Figure 1 | The overall architecture of PGMG.** (a) The construction of pharmacophore networks. We use the shortest paths on the molecular graph to approximate the Euclidean distances between two pharmacophore features and construct a fully connected graph to represent a pharmacophore model. (b) The preprocessing of SMILES. We randomize a given canonical SMILES and corrupt it using the infilling scheme. (c) Model structure and pipelines for training and inferencing. $c$ represents the embedding vector sequences for the given pharmacophore model; $x$ represents the embedding sequence of input SMILES; z represents the latent variables for a molecule. Transformer encoder and decoder blocks are stacked with N layers. $\oplus$ denotes concatenation of two vectors and $\otimes$ denotes matrix multiplication. The overlap between the training and inferencing process is highlighted in the right panel.

To train PGMG, we need only a number of SMILES strings with no additional information attached. A training sample can be constructed using the SMILES representation of a molecule. First, the chemical features of a molecule are identified using RDKit[27] and we randomly select some of them to build a pharmacophore network $G_p$. As shown in **Figure 1a**, we approximate the Euclidean distance in the three-dimensional Euclidean space in a pharamacophore using the length of the shortest path between two pharmacophore features on the molecular graph. The analysis of the two distances can be found in **Figure S1.** Next, a molecule is represented as a randomlized SMILES string and then segmented into a token sequence $s$. We then corrupt $s$ to get the encoder input $s'$ by using the infilling scheme[28] and obtain a training sample $(G_p, s, s')$. Since we avoid the use of target-specific active data in the training stage, PGMG bypasses the problem of data scarcity on active molecules.

When using the trained model to generate molecules, a pharmacophore model is required. The generation process is as follows. Given a pharmacophore model $c$, a set of latent variables $z$ is sampled from the prior distribution $p(z|c)$, which in our case is the standard Gaussian distribution $N(0, I)$, and molecules are then generated from the conditional distribution $p(x|z, c)$. There are multiple ways to construct a pharmacophore model using various active data types and this is where the flexibility of the PGMG approach comes in. We employ both ligand-based and structure-based approaches to build pharmacophores and use them to generate active molecules for de novo drug design.

**Performance of PGMG on the unconditional molecule generation task.**

We evaluate our model's performance on the unconditional molecule generation task by comparing it with other SMILES-based methods including ORGAN[29], VAE[4], SMILES LSTM[5], and Syntalinker[16]. We train PGMG and other SMILES-based models on the ChEMBL dataset[30] based on the train-test split used in the GuacaMol benchmark[31]. Since PGMG is a conditional model, we approximate the unconditional distribution by generating molecules based on randomly sampled pharmacophore features. The molecule generation performance is evaluated by four metrics including validity, novelty, uniqueness, and ratio of available molecules (see Methods for the definition of the metrics). The comparison results of PGMG and four other SMILES-based methods on the four metrics are shown in Table 1. The results of ORGAN, VAE, SMILES LSTM on validity, novelty and are taken from the GuacaMol benchmark directly.

Table 1 Performance of PGMG and SMILES-based models on ChEMBL.

| Methods | Validity | Uniqueness | Novelty | Ratio of available molecules |
|---|---|---|---|---|
| ORGAN | 0.379 | 0.841 | 0.687 | 21.9 % |
| VAE | 0.870 | 0.999 | 0.974 | 84.7 % |
| SMILES LSTM | 0.959 | **1.000** | 0.912 | 87.5 % |
| Syntalinker | **1.000** | 0.880 | 0.903 | 79.5 % |
| PGMG | 0.982 | 0.979 | **0.976** | **93.8 %** |

As shows in **Table 1**, PGMG performs better in novelty and the ratio of available molecules, while keeping the same level of validity and uniqueness as the top models. The ratio of available molecules is the ratio of unique novel valid molecules to all generated molecules, and equals product of the previous three metrics, as a composite metric to assess the performance of the model to generate novel molecules. PGMG achieves the highest the ratio of available molecules. Comparing to the second-best method, PGMG improves the ratio of available molecules by 6.3%. Among the SMILES-based methods, SMILES LSTM[5] performs the best in uniqueness, while Syntalinker[16] performs the best in validity.

To test whether PGMG catches the distribution of training dataset, we further examine the physicochemical properties of the generated molecules. The distribution of physicochemical properties of the generated molecules and the molecules in the training dataset are compared in **Figure 2**. We find that the structural properties distribution such as the topological polar surface area (TPSA), the number of rotatable covalent bonds, and the molecular weight of the generated molecules are generally consistent with the training set, and the physicochemical properties such as QED, LogP and SA are close to the training set distribution. This demonstrates that PGMG captures the distribution of molecules in the training dataset well.

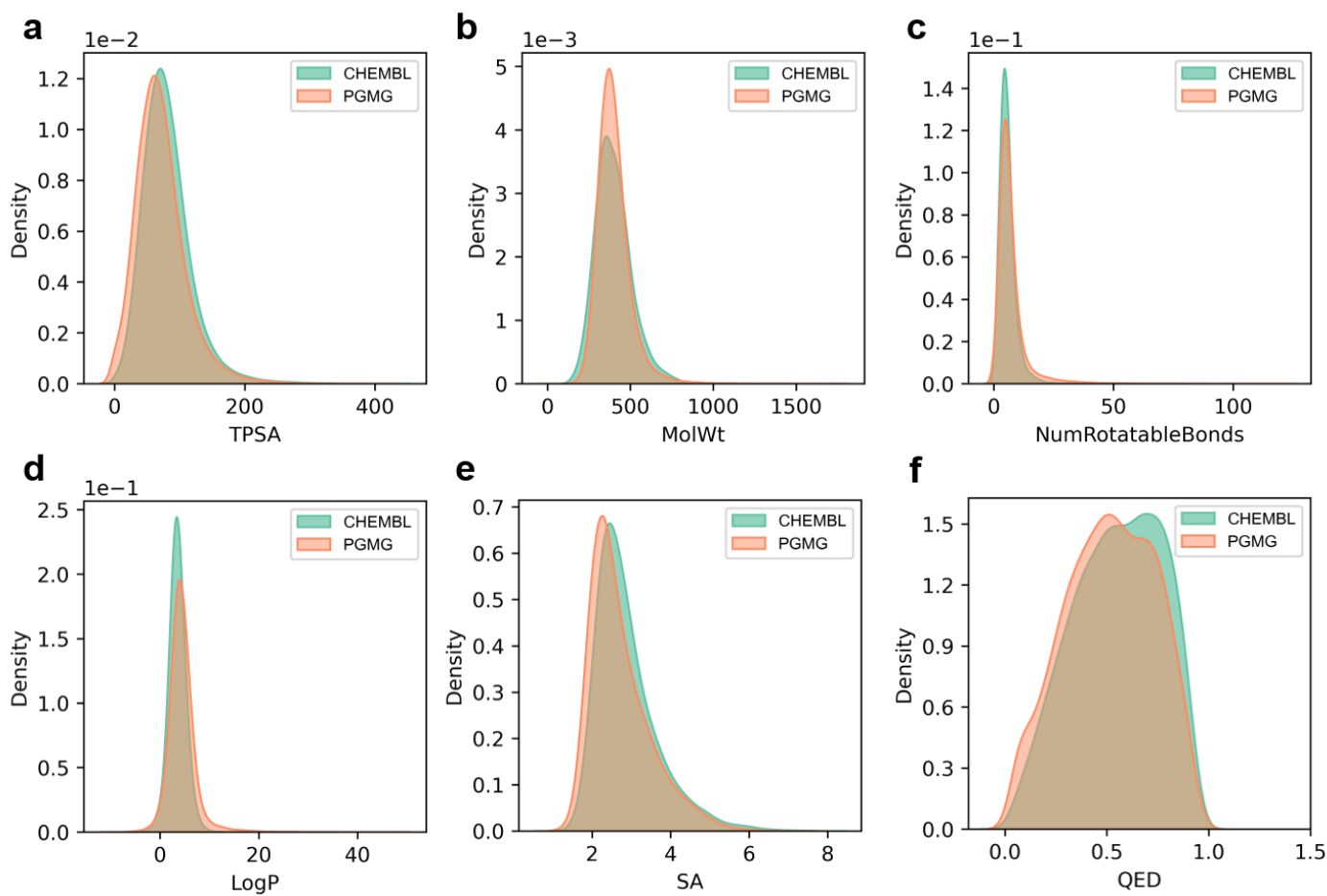

**Figure 2 | Distribution of chemical properties for the ChEMBL training set and the molecules generated by PGMG.** The scientific notation at the upper left of the figure indicates the scaling of the vertical coordinates.

**PGMG can generate bioactive molecules satisfying given pharmacophores.**

We evaluate the extent to which the generated molecules fit the given pharmacophore models and predict binding affinity between protein receptors and molecules generated using PGMG through the molecular docking tool vina[32]. We use a match score to estimate the matching degree between each molecule-pharmacophore pair (see calculation of match score section of the Supplementary Information for details).

We extract a random pharmacophore model from each molecule in the test dataset. About 230,000 molecules in total are generated from those random pharmacophore models and the match score is calculated between each pair. For comparison, we also calculate the match score between 230,000 random molecules from the ChEMBL dataset[30] and the selected pharmacophores. The result is shown in **Figure 3a**.

As can be seen from **Figure 3a**, 86.3% of the generated molecules have matching scores concentrated in the range of 0.8-1.0, with 77.9% having a matching score of 1.0. Meanwhile, the matching degrees for the random molecules are centered at 0.45, with only 4.8% having a matching score of 1.0. This result demonstrates PGMG's ability to generate molecules satisfying the given pharmacophore models.

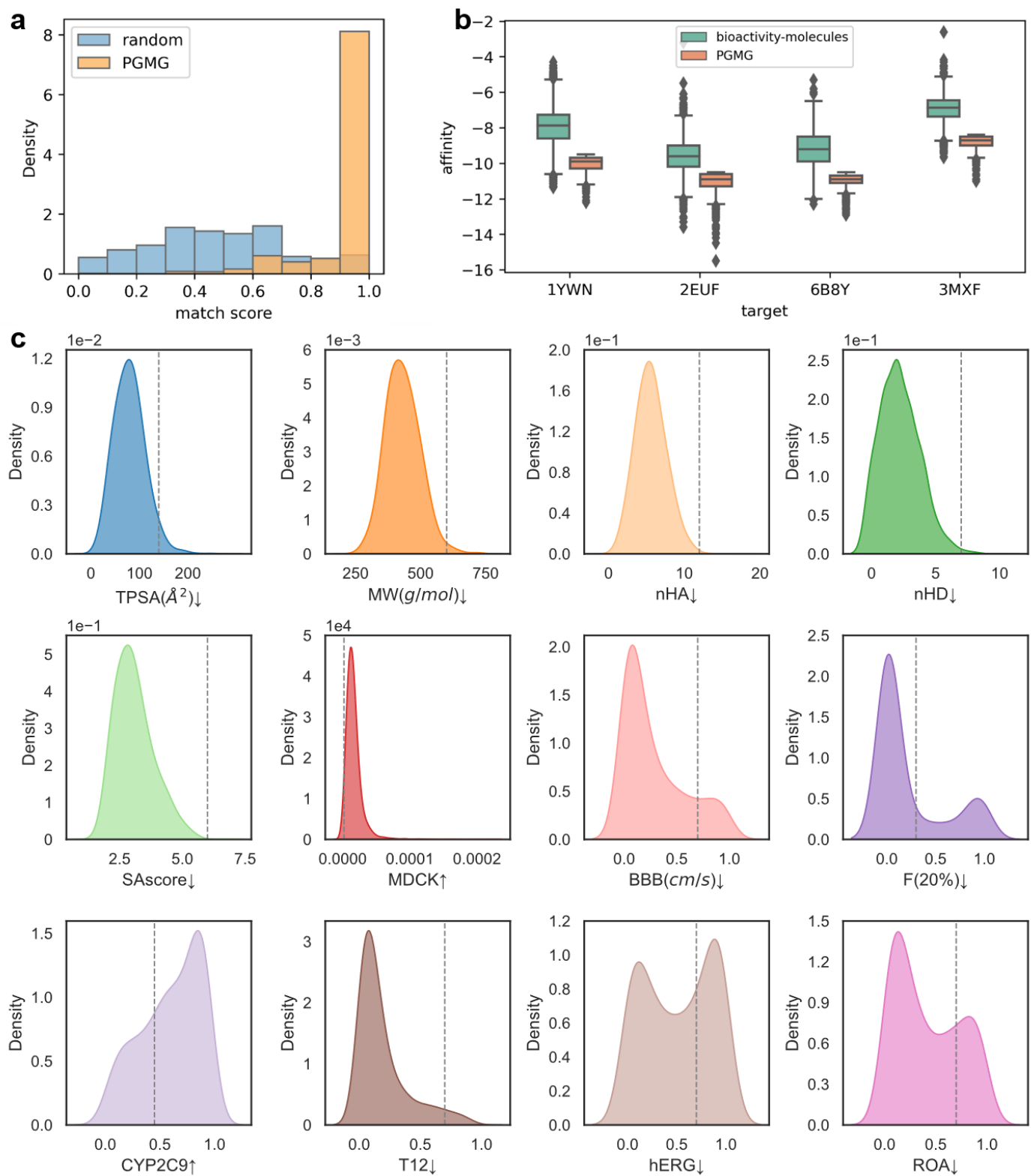

**Figure 3 | Pharmacophore matching test results and the distribution of four target docking scores.** (a) The match score of random selected molecules and PGMG generated molecules. (b) The distributions of the predicted affinity of top 1000 molecules generated by PGMG over VEGFR2 (PDB: 1YWN), CDK6 (PDB: 2EUF), TGFβ 1 (PDB: 6B8Y), BRD4 (PDB: 3MXF), and the affinity for the known bioactivity molecules corresponding to these targets. (c) Distributions of ADMET properties are calculated using ADMETlab 2.0[33] of top 1000 molecules generated by PGMG. The threshold of each property according to ADMETlab 2.0 is given as the dashed line. "↑" indicates that the distribution greater than the threshold satisfies the expected property, while "↓" indicates that the part of lower than the threshold satisfies the expected property. TPSA represents the topological polar surface area, optimal: 0~140 (Å$^2$); MW denotes Molecular Weight, Optimal:100~600; nHA represents the number of hydrogen bond donors, optimal: 0~7; nHD represents the number of hydrogen bond acceptors, optimal: 0~12; SAscore represents the synthetic

accessibility score, optimal: 0~6; Madin−Darby Canine Kidney cells (MDCK) measure the uptake efficiency of a drug into the body, optimal: >2 x $10^{-6}$ (cm/s); BBB measures the ability of a drug to cross the blood-brain barrier to its molecular targets, qualified value: 0-0.7; F(20%) denotes human oral bioavailability 20% which assess the efficiency of the drug delivery to the systemic circulation, optimal: 0~0.3; CYP2C9 assess drug metabolism reactions, the closer to 1, the more likely it is to be an inhibitor; T12 represents the half-life of the drug, qualified value: 0-0.7 and hERG evaluates whether the molecule is toxic to the heart, qualified value: 0-0.7; ROA measures acute toxicity in mammals, qualified value: 0-0.7. Where a molecule with a property in the optimal range means that the property is optimal, and a molecule with a property in the qualified range means that there is no obvious evidence that the property of the molecule is defective. The scientific notation at the upper left of the figure indicates the scaling of the vertical coordinates.

To further examine the binding activity of molecules generated by PGMG through the guidance of pharmacophores, we obtain pharmacophore models with known target structure from the literature[34, 35, 36, 37]. These targets include VEGFR2, CDK6, TFGβ 1, BRD4. For each pharmacophore model, 10,000 molecules are generated by PGMG. Autodock vina[32] is used to calculate the binding affinities of generated molecules. And then, we select the top 1000 molecules with the strongest binding affinity. For comparison, we acquire the known bioactivity molecules for the four targets from CHEMBL, including 13299, 1648, 1885 and 4786 bioactivity molecules, respectively. In **Figure 3b**, we show the affinity distributions of the top 1000 molecules generated by PGMG and the affinities for the known bioactivity molecules from CHEMBL. The average affinity of the top 1000 molecules generated by PGMG is -10.0 kcal/mol (1YWN), -11.1 kcal/mol (2EUF), -11.0 kcal/mol (6B8Y) and -8.8 kcal/mol (3MXF), and the average affinity of the known bioactivity molecules is -8.0 kcal/mol (1YWN), -9.6 kcal/mol (2EUF), -9.2 kcal/mol (6B8Y) and -7.0 kcal/mol (3MXF) respectively. The distribution of affinities suggests that PGMG can generate desired bioactive molecules.

To evaluate if PGMG is capable of generating drug-like molecules, we further calculate the pharmacokinetics properties (absorption, distribution, metabolism, excretion) and toxicity (ADMET) of the top 1000 molecules. The ADMET distributions of the top 1000 molecules are illustrated in **Figure 3c.** Most of the molecules generated by PGMG satisfy the pharmacokinetic properties and toxicity constraint for drug candidate according to the standard proposed by ADMETlab 2.0[33]. And the majority of the generated molecules are predicted with no obvious toxicity to the heart.

**Structure-based drug design**

Structure-based drug design is a powerful drug design strategy to generate the desired bioactivity molecules using the structure of specific target[38]. We use four targets (VEGFR2, CDK6, TFGβ 1, BRD4) from the above section with pharmacophore models which are built using ligand-receptor complex as examples to further demonstrate the performance of PGMG in structure-based drug design. It should be noted that the construction of pharmacophore models does not necessarily need any ligand. We choose these pharmacophore models for the convenience of the following analyses. We compare several top affinity conformations of the generated molecules with the top affinity conformation of the reference ligand in the crystal complex. **Figure 4** shows the binding sites of the four receptors with corresponding molecules. Most of the generated molecules share the same amino acid residues as the reference ligand, which indicates that those generated molecules are capable of fitting into the binding site as well as the reference one.

In **Figure 4(a-c)**, the generated molecules of 1YWM have a similar structure with the reference. As for 2EUF and 6B8Y, despite the structural differences between the generated molecules and the reference molecule, the generated molecules (**Figure 4 (e-g, i-k)**) share some common important functional groups as the reference ligands (**Figure 4h, Figure 4l**). And interestingly, we find that the structures of the generated molecules may differ from the reference ligand (**Figure 4p**) in a good way. For example, the molecules generated by PGMG for 3MXF (**Figure 4 (m-o)**) can bind to D88, P86, and P82 amino acid residues other than N140 (**Figure 4p**). This finding suggests that PGMG may have the potential in exploring new binding sites. Besides, we exam the drug-likeness using SA and hERG. SA is designed to estimate the ease of synthesis of drug-like molecules, and it's easy to synthesize when SA is less than 6. The hERG is a toxicity metric. Abnormal hERG values for a drug may lead to palpitations, syncope, and even sudden death. This metric measures the probability that a molecule will be toxic. Empirically, over 0.7, the molecule is considered toxic. These generated molecules perform well on SA and hERG. The above results show that PGMG can design molecules that not only fit well into the binding site but also exhibit drug-like quality in the structure-based drug design.

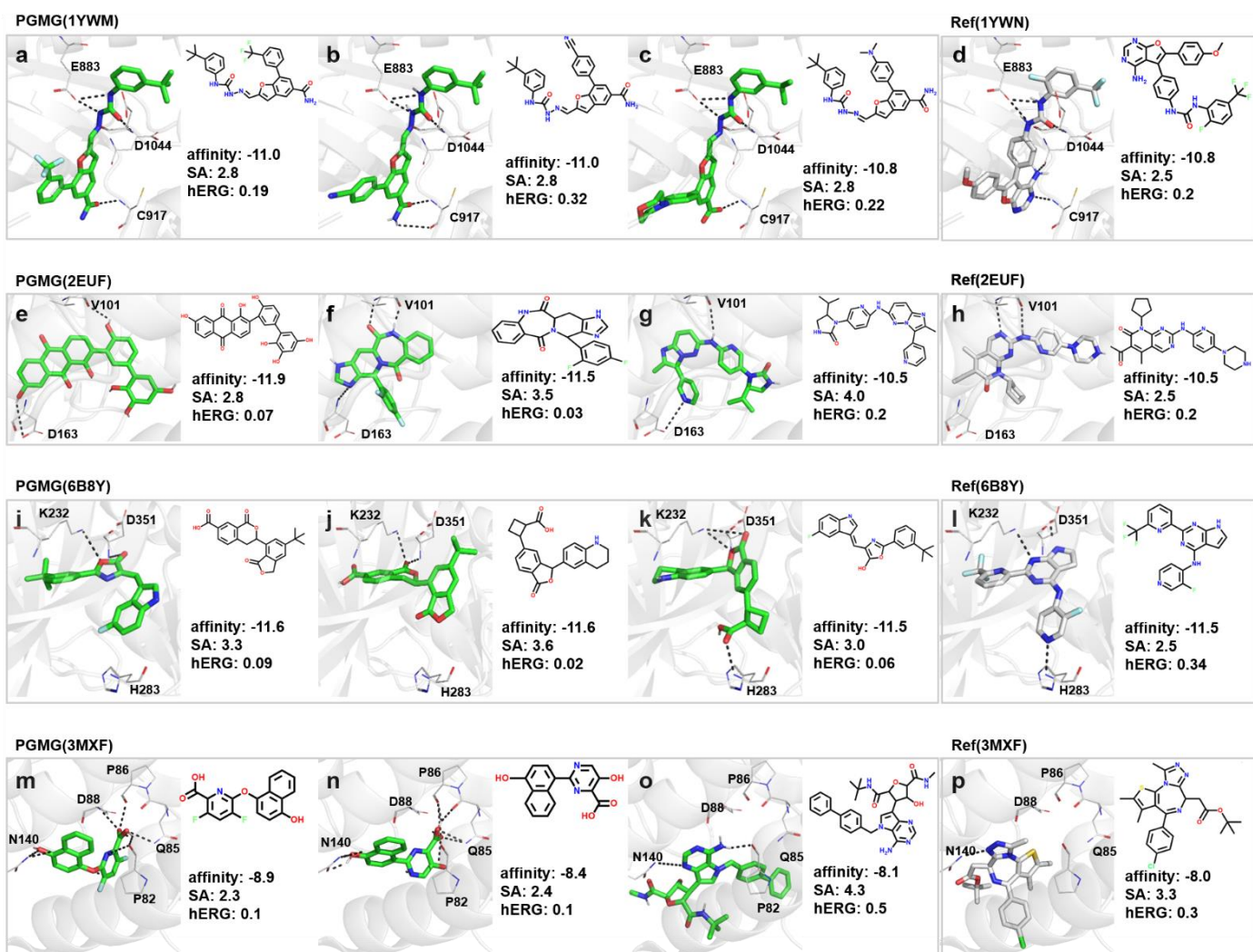

**Figure 4 | A display of the binding sites of the molecules generated by PGMG in structure-based drug design.**

**Ligand-based drug design**

Although structure-based drug design is a successful and highly attractive strategy, there are some prerequisites to use this strategy, including a certain target, a high-resolution crystal structure of the target, and some identified interaction sites. However, it is not easy to reach the above prerequisites. Ligand-based drug design is capable of designing drug molecules based on the conformational superposition of known active molecules when the target is unknown or the binding site is unclear. And it has been widely used in drug discovery, such as the search for new drugs for drug resistance. Squalene oxidase is the target for ringworm, superficial skin fungal infections, and other diseases. Butenafine and terbinafine are typical inhibitors for squalene oxidase[39]. However, these inhibitors are prone to drug resistance and side effects including skin erythema, burning, and itching. Therefore, it is urgent to design novel squalene oxidase inhibitors. Here, we generate 200 molecules using a pharmacophore model constructed from squalene oxidase inhibitors.

As shows in **Figure 5**, the generated molecules align well to the active conformation of terbinafine which is obtained from drugbank[40]. The listed molecules match well with the desired pharmacophore features, including two hydrophobic centers, a positive charge center, and an aromatic ring center. Here, we notice that PGMG has a good grasp of the equivalence of different substructures under the same pharmacophore feature. It matches the aromatic center with pyrrole, thiophene, and pyrimidine, and the hydrophobic center with aliphatic, cycloalkane, and benzene. This result shows that PGMG can generate diverse molecules while maintaining the important properties of the substructures the same as the known inhibitor.

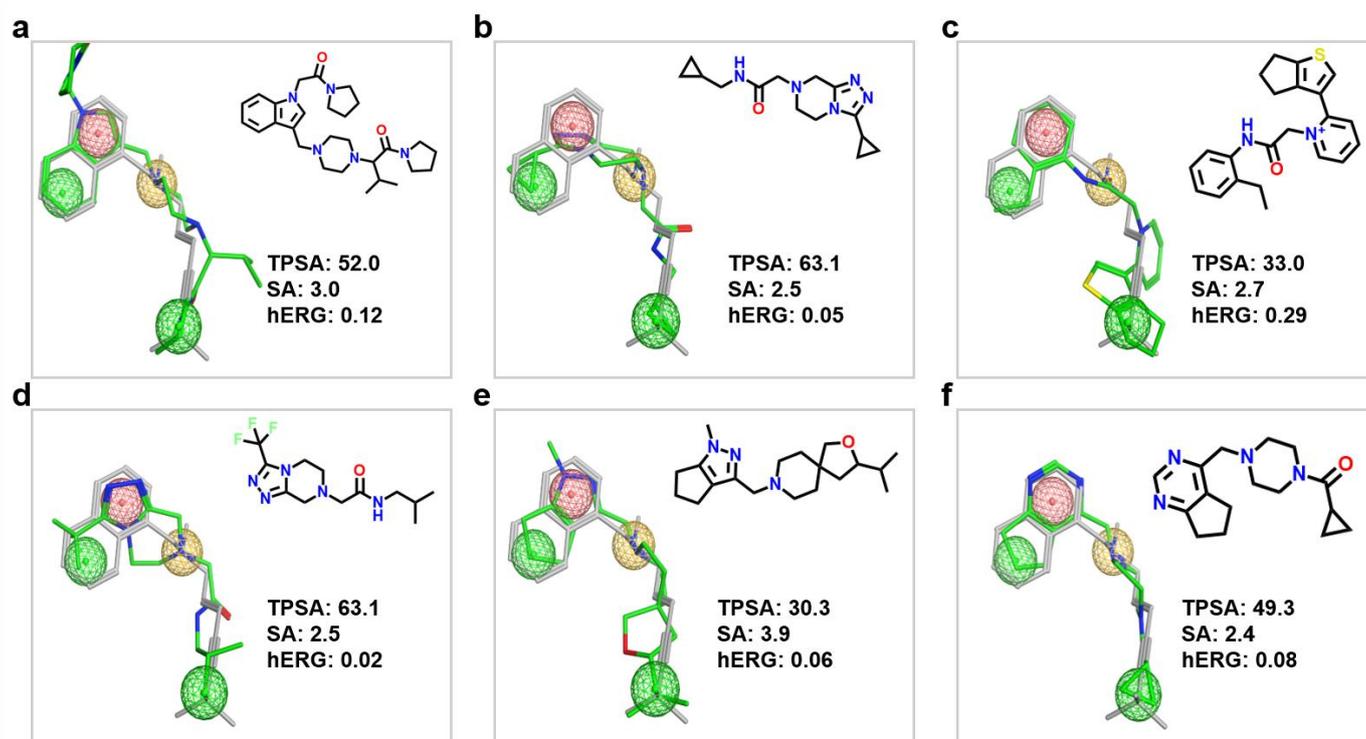

**Figure 5 Alignment diagrams of terbinafine, pharmacophore model, and molecules generated by PGMG.** The different colored spheres represent different pharmacophore features. Aromatic center is red, the positive charge center is yellow, and hydrophobic centers are green. The grey molecules represent terbinafine, and the green molecules represent the molecules generated by PGMG.

To further assess the pharmacokinetics and toxicity of the generated molecules, we calculate the TSPA, SA, and hERG of the generated molecules. See the previous section for a detailed SA and hERG description. TSPA is a molecular descriptor measuring drug transport properties such as intestinal absorption and blood-brain barrier (BBB) penetration. The TPSA in the range of 0-140 means optimal. Of the six molecules generated by PGMG, their TSPA, SA, and hERG values are within the rational range. From Figure 5, we can see that PGMG is able to generate molecules that match the pharmacophore model and meet the overall criteria for TSPA, SA, and hERG.

**Lead compound optimization**

Lead optimization refers to the improvement of one or more properties of a hit compound by chemical modification. The optimization objectives include adjusting the molecular flexibility ratio, improving the pharmacokinetic properties, or enhancing the bioavailability. Here we show how PGMG can help with lead compound optimization using Lavendustin A as a case study. Lavendustin A is an inhibitor of epidermal growth factor receptor (EGFR), while the lipophilicity of Lavendustin A is too poor to cross the cell membrane. It has been shown that improving the lipophilicity of Lavendustin A can lead to nanomolar levels of IC50 inhibition activity at the cellular level[41]. Therefore, we construct Lavendustin A's pharmacophore and aim to improve the lipophilicity of Lavendustin A by PGMG.

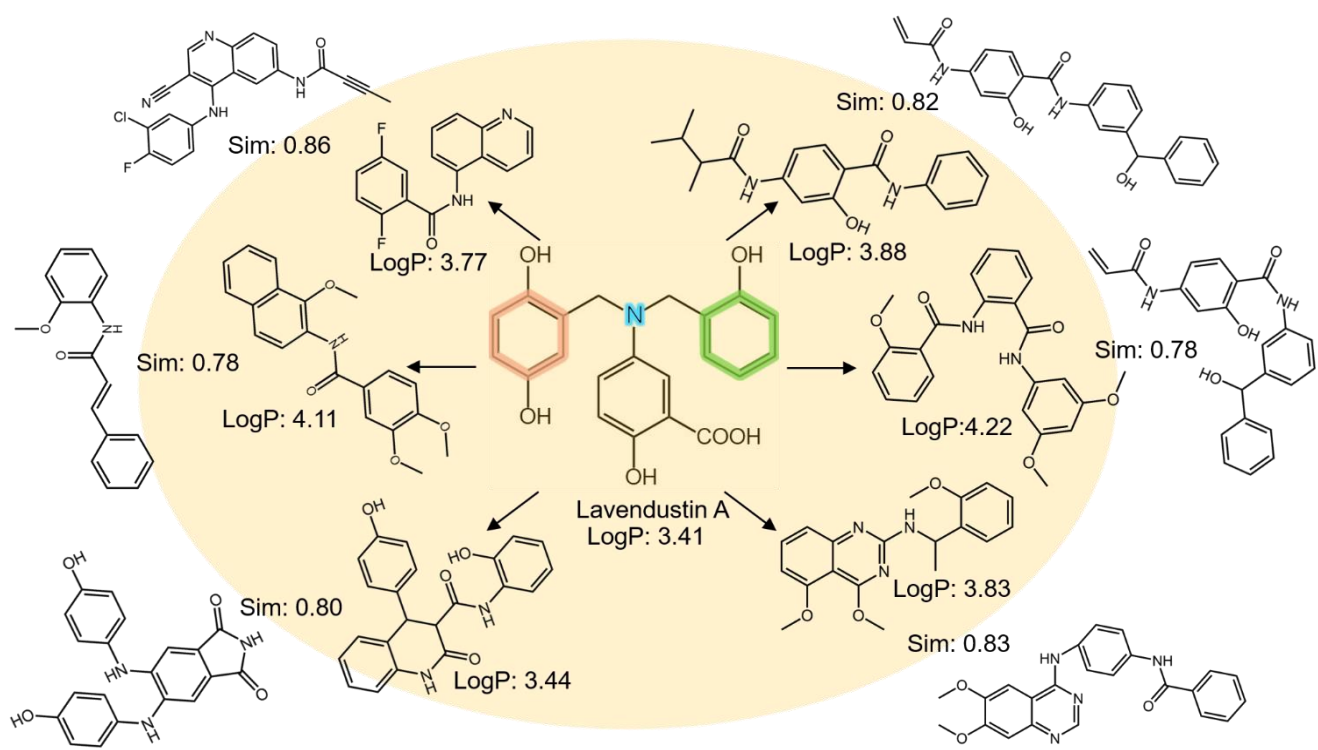

**Figure 6 | Display diagram of the molecule generated by PGMG with known inhibitors in the case of Lavendustin A optimization.** Molecules generated by PGMG are shown inside the circle and their closest active nearest neighbors are shown outside the circle. The colors indicate the pharmacophore features extracted from Lavendustin A. Red corresponds to the aromatic center, blue represent the hydrogen-bonded acceptor, and green represent the hydrophobic center.

We filter the generated molecules with lipophilicity (LogP > 3.41) to obtain 400 molecules with a higher lipophilicity than Lavendustin A. We calculate Tanimoto similarity using MACCSkeys Fingerprints with RDKit[27] between the obtained molecules and the 1500 EGFR inhibitors acquired from the ExCAPE database[42]. **Figure 6** shows some examples of the generated molecules with their closest EGFR inhibitors obtained from the ExCAPE database and their respective Tanimoto similarities. We find that the generated molecules have high similarity to the EGFR target active molecules in the ExCAPE database, which are not included in the training set. And they all own the three pharmacophore features of the aromatic ring, hydrogen-bonded acceptor, and hydrophobic center. Based on the assumption that structurally similar molecules have similar properties, the similarity result demonstrates that molecules generated by PGMG have a probability of inhibiting EGFR. To some extent, the generated molecules gain structural diversity while maintaining the consistency of the pharmacophore. Overall, PGMG can optimize certain properties and maintain the bioactivity of a given lead compound.

**Discussion**

In this work, we develop a pharmacophore-guided deep learning approach for bioactive molecule generation called PGMG. We manage to use pharmacophores as the only constraint during the generation process by (1) encoding both pharmacophore features and spatial information of a given pharmacophore into a complete graph with node and edge attributes and (2) introduce latent variables so that a molecule can be uniquely characterized by a pharmacophore and a set of latent variables to handle the many-to-many relationship of pharmacophores and molecules. Our approach offers some advantages over current molecule generation methods. Firstly, PGMG provides a way to utilize different types of activity data in a uniform representation, allowing it to overcome the problem of data scarcity and be used in various situations. Secondly, pharmacophores are biologically meaningful and can incorporate biochemists' knowledge, which provides a strong prior and certain interpretability into the generation process. Lastly, after training, PGMG can be directly applied to different targets without further fine-tuning. Besides, it is also worth mentioning that the training scheme itself does not require any activity data to proceed. This training scheme may be useful for other generative methods.

PGMG makes solid progress on the challenging problem of generating desired bioactivate molecules in various scenarios when known active data is scarce. When a target structure is available, PGMG is competent to design a large number of molecules that bind affinity better than the specific-target active molecules obtained from the ChEMBL database. Given the pharmacophore for certain targets, the PGMG can also be utilized to design dual or multi-target molecules. Besides, we expect that PGMG can be adopted to prepare chemical libraries to replace those used in HTS campaigns to improve virtual screen efficiency, as it can provide a sufficient number of candidate drug-like molecules for a specified target. Then, this method performs well in ligand-based drug design, which has wide use in drug design when the target structure is absent. The ligand-based case shows that PGMG is able to generate high-quality bioactivity molecules that match the pharmacophore model with structural diversity. This result implies that PGMG can be applied to multiple drug design scenarios such as researching alternative medicine, drug resistance, and scaffold hopping. Finally, the case of lead optimization demonstrates that PGMG can optimize the molecule's properties while maintaining the bioactivity and scaffold diversity of the generated molecules. The results demonstrate that PGMG is a promising approach for structure-based drug design, high-throughput screening, ligand based-drug design, and lead optimization in a real drug discovery setting.

We believe that the de novo drug design is a complicated and situation-specific problem, and computational methods should try to get more assistance from chemists' experience and judgements. PGMG benefits from this idea a lot. Some limitations of PGMG should be acknowledged. PGMG currently does not support exclusion volume in pharmacophore models and as we focus on the task of generating molecules with desired activity, PGMG does not explicitly constrain the properties of the generated molecules. A future direction of our work is to include the exclusion volume and other features into PGMG and make the generated molecules to be more controllable and malleable. And furthermore, designing multi-conditional generation models to generate active molecules with specified properties is the ultimate goal of drug design, we will continue to work towards this objective.

## Methods

### Datasets

We use the ChEMBL 24 dataset containing more than 1.25 million molecules to train PGMG. ChEMBL is a collection of bioactivity data for various targets and compounds from the literature. It contains 13 types of atoms (T = 13): H, B, C, N, O, F, Si, P, S, Cl, Se, Br, and I. Each bond is either a no-bond, single, double, triple or aromatic bond (R = 5).

We also use the ZINC[43] molecule dataset from JTVAE[7] for our ablation study. It contains 220,000 molecules in the training data, 11 types of atoms (T = 11): H, B, C, N, O, F, P, S, Cl, Br, and I. Each bond is either a no-bond, single, double, triple or aromatic bond (R = 5).

The structure of four targets VEGFR2, CDK6, TFGβ 1, BRD4 are downloaded from PDB[44] database.

### Representation of Pharmacophores and Molecules

A pharmacophore model consists of several chemical features and their spatial descriptions and are represented by a fully connected graph with chemical feature types as node attributes and distances as edge weights (a detailed description of the pharmacophore graph and the preparation of the pharmacophore graph is included in SI). We apply a state-of-the-art graph neural network, Gated-GCN[25], to embed the graph with consideration of node attributes and edge attributes.

Molecules are represented in SMILES format. Symbols of stereochemistry like '@' '/' are removed because stereochemistry information does not exist in the graph representation of a pharmacophore and it is not difficult to list all stereoisomers of a molecule. Then the SMILES string is separated into a sequence of tokens corresponding to heavy atoms and structural punctuation marks. For example, the SMILES string *C(C[NH2-])OC(=O)Cl* will be split to *C ( C [NH2-] ) O C ( = O ) Cl*, where each token will be embedded into a vector.

### Encoder and Decoder

An illustration of the encoder and decoder networks can be found in Figure.1. The encoder and decoder network are adapted from the standard transformer[26] architecture with each consisting of several layers of stacked transformer encoder block and transformer decoder block. The difference between the transformer encoder and decoder blocks is that the encoder block uses only self-attention modules and the decoder block uses cross-attention modules to incorporate context in the generation process. Some modifications are made to handle our inputs and to better suit the variational autoencoder structure of PGMG.

We first calculate the latent variables $z$ of molecule $x$ given pharmacophore $c$ by the encoder network. The encoder input is a concatenation of molecule and pharmacophore features. Following BART[28], positional and segment encoding is added to the input sequence:

$$Input_{encoder} = (E'_p; E'_m) \quad (2)$$
$$E'_{m_i} = E_{m_i} + SE_m + PE_i \quad (3)$$
$$E'_{p_j} = E_{p_j} + SE_p \quad (4)$$

where $Input_{encoder}$ is the input representation, $E_{p_j}$ is the j-th pharmacophore feature vector, $E_{m_i}$ is the i-th token embedding of molecule features, $SE_m$ and $SE_p$ are two segment embedding vectors for molecule features and pharmacophore features, and $PE_i$ is the positional embedding for i-th token. After several layers of transformer encoder block, the molecule features are averaged by an attention pooling layer to obtain the final molecule representation $h_x$. $h_x$ is then fed into two separate sub-networks to compute the mean $\mu$ and log variance $\log \Sigma$ of the posterior variational approximation. Latent variables $z$ are then sampled from the Normal distribution $N(\mu, \Sigma)$.

The decoder network takes the latent variables $z$ and pharmacophore features as input:

$$input_{decoder} = (E'_p; E'_z) \quad (5)$$
$$E'_{z_i} = z_i + SE_z + PE_i \quad (6)$$

where $E'_p$ is the same as described above, $SE_z$ is the segment embedding for latent variables, and $PE_i$ is the positional embedding for i-th token. The decoder then uses $input_{decoder}$ to generate target SMILES autoregressively. Each token is determined on the basis of previously generated tokens and context:

$$o_i = (argmax)_{o_i} P(o_i | o_{<i}, c, z) \quad (7)$$

where $o_i$ is i-th generated token.

**Loss Function**

PGMG's model is trained in an end-to-end manner. The Loss function consists of three different terms, KL Loss, LM Loss, and the mapping loss. The first two terms are the negative evidence lower bound (ELBO) of the log likelihood $\log P_\theta(x|c)$:

$$\log P_\theta(x|c_p) = \log \int P_\theta(x|c,z) P_\phi(z|c) dz$$
$$\geq -KL(P_\phi(z|x,c) || P_\theta(z|c)) + E_{P_\phi(z|x,c)}[\log P_\theta(x|z,c)] \quad (A)$$
$$\approx -KL(P_\phi(z|x,c) || P_\theta(z|c)) + \log P_\theta(x|z,c) \quad (B) \quad (8)$$

where $KL$ denotes the Kullback-Leibler divergence and we assume $P_\theta(z|c)$ the prior distribution of $z$ to be a standard gaussian $N(0, I)$. We call the left part of (A) KL Loss and it serves as a regulation term to mitigate the gap between the true prior distribution of $z$ and the posterior distribution and to make the latent space of $z$ smoother. The expectation term on the right part of (A) is estimated through sampling, and it is optimized using Monte Carlo estimation with one data point for each sample[45]. This gives us the right part of (B). Since $m$ takes form of the SMILES string, we call it the language modeling loss (LM Loss).

The third part of PGMG's loss function is the mapping loss. It evaluates the model's performance in predicting the mapping between heavy atoms and pharmacophore elements. We use the mapping loss as a regulation term to help alleviate the problem of posterior collapse. The mapping score of the $i^{th}$ pharmacophore $p_i$ and the $j^{th}$ output token $o_j$ is calculated as

$$mapping_{score_{ij}} = \sigma\left(g(W_p E_{p_i}) \odot g(W_o E_{o_j})\right) \quad (9)$$

where $E_{p_i}$ and $E_{o_j}$ are the embedding vectors of $p_i$ and $o_j$ respectively, $W_p$ and $W_o$ are two learnable matrices to project two different embeddings into the same space, $\odot$ is the dot product, $\sigma$ is the sigmoid function, and $g$ is the ReLU function. The calculation of mapping scores can be vectorized as

$$mapping_{score} = \sigma\left(g(W_p E_p)\, g(W_o E_o)\right) \quad (10)$$

Since SMILES format contains tokens other than atom symbols, we mask them when calculating the mapping loss. The mapping loss is then calculated as the cross-entropy of the masked scores and labels. An illustration of the masked mapping score and label is given in Supplementary **Figure S3**.

**Training details and model parameter settings**

During training, we inject noise into the input to make training more robust by using the infilling scheme. Some random subsequences in every input sequence are replaced with a single *[mask]* token. Teacher forcing technique is applied to the generation process during training, by which we replace the previously generated tokens with the ground truth to produce the next token. Aside from the mapping loss introduced before, another approach we use to alleviate posterior collapse is KL annealing[46], where an increasing coefficient is used to control the size of KL Loss.

We use the same model parameters in both ChEMBL and ZINC datasets. The hidden dimension is 384. The transformer encoder blocks and transformer decoder blocks are stacked 8 times. We use an 8-head attention and the feed-forward dimension is 1024. We use Adam optimizer to train the model with a 3e-4 learning rate and a 1e-6 weight decay rate. Cosine learning rate annealing is applied with a cycle length of 4 epochs. We use the gradient clipping technique and set the maximum gradient to be 5. Since the ChEMBL dataset contains a lot more molecules compared to the ZINC dataset, it requires less training epochs to reach a similar validation performance. Thus, the number of training epochs for the former is 32 and 48 for the latter.

**Evaluation**

Firstly, four different metrics on 2D level, validity, uniqueness, novelty, and ratio of available molecules are employed to evaluate the ability of the PGMG to generate novel molecules. Validity is the percentage of chemically valid molecules with generated SMILES. Uniqueness measures how many valid molecules are non-repetitive. Novelty refers to the percentage of generated chemically valid molecules not present in the training set. And the ratio of available molecules is the proportion of novel molecules in all generated results. These metrics are calculated as follows:

$$validity = \frac{\#Number\ of\ chemically\ valid\ SMILES}{\#of\ generated\ SMILES} \quad (11)$$

$$uniqueness = \frac{\#of\ non-duplicate, valid\ SMILES}{\#of\ valid\ SMILES} \quad (12)$$

$$novelty = \frac{\#of\ novelty\ molcules\ not\ in\ training\ set}{\#of\ unique\ molecules} \quad (13)$$

$$ratio\ of\ available\ molecules = \frac{\#of\ novel\ molecules}{\#of\ generated\ SMILES} \quad (14)$$

Secondly, goal-directed metrics are evaluated by the match score, which indicates the match degree of the generated molecules to the specified pharmacophore (see calculation of match score section of the Supplementary Information for details). We further evaluate the binding activity of the generated molecules to the target using affinity calculated by Autodock vina[32]. Finally, we use ADMETlab 2.0[33] to predict the ADMET properties of the generated molecules and to assess the drug-like potential of the generated molecules.

**Acknowledgements**

This work is financially supported by the National Natural Science Foundation of China under Grants (No. 61832019 to M.L.), Hunan Provincial Science and Technology Program (2019CB1007) [M.L.], and European Research Council (No. 716063 to J.T.)

**Author Contributions**

M.L. and J.T. guided the research and provided the experimental platform. H.Z., R.Z, and M.L conceived the initial idea and started the project. H.Z collected and preprocessed the data and R.Z designed the model. R.Z performed the generation experiments and H.Z performed the case studies. H.Z, R.Z, J.T., and M.L wrote the paper.

**Declaration of Interests**

The authors declare no competing interests.

**Data availability**

The data that support the findings of this study are available at https://github.com/CSUBioGroup/PGMG.

**Code availability**

The code used to generate results shown in this study is available at https://github.com/CSUBioGroup/PGMG.

# reference

1. Lipinski CA, Lombardo F, Dominy BW, Feeney PJ. Experimental and computational approaches to estimate solubility and permeability in drug discovery and development settings. *Advanced drug delivery reviews* **23**, 3-25 (1997).
2. Bohacek RS, McMartin C, Guida WC. The art and practice of structure-based drug design: a molecular modeling perspective. *Medicinal research reviews* **16**, 3-50 (1996).

3. Goodnow Jr RA. Hit and lead identification: Integrated technology-based approaches. *Drug Discovery Today: Technologies* **3**, 367-375 (2006).
4. Gómez-Bombarelli R*, et al.* Automatic chemical design using a data-driven continuous representation of molecules. *ACS central science* **4**, 268-276 (2018).
5. Segler MH, Kogej T, Tyrchan C, Waller MP. Generating focused molecule libraries for drug discovery with recurrent neural networks. *ACS central science* **4**, 120-131 (2018).
6. Mahmood O, Mansimov E, Bonneau R, Cho K. Masked graph modeling for molecule generation. *Nature communications* **12**, 1-12 (2021).
7. Jin W, Barzilay R, Jaakkola T. Junction Tree Variational Autoencoder for Molecular Graph Generation. In: *Proceedings of the 35th International Conference on Machine Learning* (eds Jennifer D, Andreas K). *PMLR*, 2323-2332 (2018).
8. Maziarka Ł, Pocha A, Kaczmarczyk J, Rataj K, Danel T, Warchoł M. Mol-CycleGAN: a generative model for molecular optimization. *Journal of Cheminformatics* **12**, 1-18 (2020).
9. Zhou Z, Kearnes S, Li L, Zare RN, Riley P. Optimization of molecules via deep reinforcement learning. *Scientific reports* **9**, 1-10 (2019).
10. Tkatchenko A. Machine learning for chemical discovery. *Nat Commun* **11**, (2020).
11. Elton DC, Boukouvalas Z, Fuge MD, Chung PW. Deep learning for molecular design—a review of the state of the art. *Molecular Systems Design & Engineering* **4**, 828-849 (2019).
12. Walters WP, Murcko M. Assessing the impact of generative AI on medicinal chemistry. *Nature biotechnology* **38**, 143-145 (2020).
13. Imrie F, Hadfield TE, Bradley AR, Deane CM. Deep generative design with 3D pharmacophoric constraints. *Chemical science* **12**, 14577-14589 (2021).
14. Méndez-Lucio O, Baillif B, Clevert D-A, Rouquié D, Wichard J. De novo generation of hit-like molecules from gene expression signatures using artificial intelligence. *Nature communications* **11**, 1-10 (2020).
15. Imrie F, Bradley AR, van der Schaar M, Deane CM. Deep generative models for 3D linker design. *Journal of chemical information and modeling* **60**, 1983-1995 (2020).
16. Yang Y, Zheng S, Su S, Zhao C, Xu J, Chen H. SyntaLinker: automatic fragment linking with deep conditional transformer neural networks. *Chemical science* **11**, 8312-8322 (2020).
17. Li Y, Pei J, Lai L. Structure-based de novo drug design using 3D deep generative models. *Chemical science* **12**, 13664-13675 (2021).
18. Luo S, Guan J, Ma J, Peng J. A 3D Generative Model for Structure-Based Drug Design. *Advances in Neural Information Processing Systems* **34**, (2021).
19. Pogány P, Arad N, Genway S, Pickett SD. De novo molecule design by translating from reduced graphs to SMILES. *Journal of chemical information and modeling* **59**, 1136-1146 (2018).
20. Skalic M, Jiménez J, Sabbadin D, De Fabritiis G. Shape-based generative modeling for de novo drug design. *Journal of chemical information and modeling* **59**, 1205-1214 (2019).
21. Schneidman-Duhovny D, Dror O, Inbar Y, Nussinov R, Wolfson HJ. PharmaGist: a webserver for ligand-based pharmacophore detection. *Nucleic acids research* **36**, W223-W228 (2008).
22. Wang X*, et al.* PharmMapper 2017 update: a web server for potential drug target identification with a comprehensive target pharmacophore database. *Nucleic acids research* **45**, W356-W360 (2017).
23. Ma Z*, et al.* Pharmacophore hybridisation and nanoscale assembly to discover self-delivering lysosomotropic new-chemical entities for cancer therapy. *Nature communications* **11**, 1-12 (2020).
24. Meslamani J, Li J, Sutter J, Stevens A, Bertrand H-O, Rognan D. Protein–ligand-based pharmacophores: generation and utility assessment in computational ligand profiling. *Journal of chemical information and modeling* **52**, 943-955 (2012).
25. Bresson X, Laurent T. Residual Gated Graph ConvNets. *Preprint at https://arxivorg/abs/171107553*, (2017).
26. Vaswani A*, et al.* Attention is all you need. *Advances in neural information processing systems* **30**,

27. Landrum G. RDKit: Open-source cheminformatics. http://www.rdkit.org.
28. Lewis M, Liu Y, Goyal N, Ghazvininejad M, Zettlemoyer L. BART: Denoising Sequence-to-Sequence Pre-training for Natural Language Generation, Translation, and Comprehension. *Preprint at http://arxivorg/abs/191013461*, (2019).
29. Guimaraes GL, Sanchez-Lengeling B, Outeiral C, Farias PLC, Aspuru-Guzik A. Objective-reinforced generative adversarial networks (ORGAN) for sequence generation models. *Preprint at http://arxivorg/abs/170510843*, (2017).
30. Mendez D, *et al.* ChEMBL: towards direct deposition of bioassay data. *Nucleic acids research* **47**, D930-D940 (2019).
31. Brown N, Fiscato M, Segler MH, Vaucher AC. GuacaMol: benchmarking models for de novo molecular design. *Journal of chemical information and modeling* **59**, 1096-1108 (2019).
32. Trott O, Olson AJ. AutoDock Vina: improving the speed and accuracy of docking with a new scoring function, efficient optimization, and multithreading. *Journal of computational chemistry* **31**, 455-461 (2010).
33. Xiong G, *et al.* ADMETlab 2.0: an integrated online platform for accurate and comprehensive predictions of ADMET properties. *Nucleic Acids Research* **49**, W5-W14 (2021).
34. Lee K, *et al.* Pharmacophore modeling and virtual screening studies for new VEGFR-2 kinase inhibitors. *European journal of medicinal chemistry* **45**, 5420-5427 (2010).
35. Shawky AM, Ibrahim NA, Abourehab MA, Abdalla AN, Gouda AM. Pharmacophore-based virtual screening, synthesis, biological evaluation, and molecular docking study of novel pyrrolizines bearing urea/thiourea moieties with potential cytotoxicity and CDK inhibitory activities. *Journal of enzyme inhibition and medicinal chemistry* **36**, 15-33 (2021).
36. Jiang J, Zhou H, Jiang Q, Sun L, Deng P. Novel transforming growth factor-beta receptor 1 antagonists through a pharmacophore-based virtual screening approach. *Molecules* **23**, 2824 (2018).
37. Yan G, *et al.* Pharmacophore-based virtual screening, molecular docking, molecular dynamics simulation, and biological evaluation for the discovery of novel BRD 4 inhibitors. *Chemical Biology & Drug Design* **91**, 478-490 (2018).
38. Pei J, Yin N, Ma X, Lai L. Systems biology brings new dimensions for structure-based drug design. *Journal of the American Chemical Society* **136**, 11556-11565 (2014).
39. Kermani F, *et al.* In vitro activities of antifungal drugs against a large collection of Trichophyton tonsurans isolated from wrestlers. *Mycoses* **63**, 1321-1330 (2020).
40. Wishart DS, *et al.* DrugBank 5.0: a major update to the DrugBank database for 2018. *Nucleic acids research* **46**, D1074-D1082 (2018).
41. Nussbaumer P, Winiski AP, Cammisuli S, Hiestand P, Weckbecker G, Stuetz A. Novel antiproliferative agents derived from lavendustin A. *Journal of medicinal chemistry* **37**, 4079-4084 (1994).
42. Sun J, *et al.* ExCAPE-DB: an integrated large scale dataset facilitating Big Data analysis in chemogenomics. *Journal of cheminformatics* **9**, 1-9 (2017).
43. Sterling T, Irwin JJ. ZINC 15–ligand discovery for everyone. *Journal of chemical information and modeling* **55**, 2324-2337 (2015).
44. Burley SK, Berman HM, Kleywegt GJ, Markley JL, Nakamura H, Velankar S. Protein Data Bank (PDB): the single global macromolecular structure archive. *Protein Crystallography* **1607**, 627-641 (2017).
45. Kingma DP, Welling M. Auto-Encoding Variational Bayes. *Preprint at http://arxivorg/abs/13126114*, (2014).
46. Bowman SR, Vilnis L, Vinyals O, Dai AM, Jozefowicz R, Bengio S. Generating sentences from a continuous space. *Preprint at http://arxivorg/abs/151106349*, (2015).

Huimin Zhu[1,†], Renyi Zhou[1,†], Jing Tang[2] and Min Li[1,*]

[1] School of Computer Science and Engineering, Central South University, Changsha 410083, China

[2] Faculty of Medicine, University of Helsinki, Helsinki, 00290, Finland

[†] These two authors contribute equally to the work.
[*] Corresponding author, limin@mail.csu.edu.cn

**Pharmacophore Description**

We use *Basefeatues.fdef* in RDKit[1], which contains a series of defined molecular substructures and their corresponding pharmacophore features, to obtain all the pharmacophore features of a molecule. The description and count for each pharmacophore feature are recorded in **Table S1**. The typical pharmacophore features include aromatic center, posionizable center (positive charge center), hydrogen-bonded acceptors, hydrogen-bonded donor, and hydrophobic center including hydrophobe, LumpedHydrophobe. As negionizable center and znbinder are rare, we label them as unknown. As shows in the **Table S1**, aromatic center, hydrophobe, hydrogen bond acceptor, hydrogen bond donor, and Lumpedhydrophobe, five kinds of pharmacophore features almost always appear in every molecule.

**Table S1. Distribution of pharmacophore features and detailed description.**

| type | description | Counts on training set | Counts on pharmacophore features | occurrence rate on training set | occurrence rate on pharmacophore |
|---|---|---|---|---|---|
| aromatic center | Number of Π-electrons conforming to the 2n+2 rule | 3148081 | 1153471 | 2.50 | 0.92 |
| posionizable center | positively charged atoms or functional groups ionized at physiological PH | 183953 | 505109 | 0.40 | 0.15 |
| hydrogen-bonded acceptor | (1) sp or sp2 hybridized oxygen atoms; (2) Sulphur atom attached to a carbon atom in a double bond; (3) Nitrogen atoms attached to carbon atoms by double or triple bonds, and imino group | 4958778 | 1768234 | 3.94 | 1.40 |
| hydrogen-bonded donor | (1) hydroxyl groups; (2) Amino, imino group | 2525490 | 880398 | 2.01 | 0.700 |
| hydrophobe | a group of continuous carbon atoms that are not connected to charged atoms or electronegative centers(non-ring) | 5235429 | 1778375 | 4.16 | 1.41 |
| Lumped-Hydrophobe | a contiguous group of carbon atoms not attached to an electrically charged atom or electronegativity center, ring | 2336036 | 833539 | 1.85 | 0.66 |
| unknown | some rare pharmacophore features, such as negionizable, znbinder | 283501 | 80726 | 0.22 | 0.064 |

Counts on the training set refer to statistics on all molecular pharmacophore features of the training set. Counts on pharmacophore features represent the number of pharmacophore features chosen at random from the training set. Occurrence rate on the training set is the average of the occurrence of a pharmacophore feature on a molecule. Occurrence rate on a pharmacophore represents the occurrence of a pharmacophore feature in the pharmacophore model.

**Molecular pre-processing**

We perform the following four steps to translate a molecule into a random pharmacophore model: 1) Use RDKit to obtain all of its pharmacophore features. 2) Select 3–7 pharmacophore features at random for each molecule; 3) Calculate the distances between the selected pharmacophore features. In this part, PGMG uses the sum of the bond lengths in the shortest path on the molecular graph between two pharmacophore features to substitute the Euclidean distances between pharmacophore features. 4) The pharmacophore model is coded as a complete graph, with the vertexes representing the pharmacophore features and the edges representing the distances between two pharmacophore features.

Table S2. Comparison of actual bond lengths with mapped bond lengths.

| Covalent bond type | bond length (nm) | Relative bond length |
|---|---|---|
| Single bond | 0.154 | 1.00 |
| Double bond | 0.134 | 0.87 |
| Aromatic bond | 0.138 | 0.91 |
| Triple Bond | 0.120 | 0.78 |

To verify the feasibility of this substitution, we select 1000 small molecules from the training set and obtain their 3D coordinates using RDKit[1]. We calculate the Euclidean distances between the pharmacophore features of these molecules. At the same time, we calculate the distances between the pharmacophore features according to the mapping rules described in **Table S2**. As illustrated in **Figure S1**, the shortest-path distances between pharmacophore features are strongly correlated with the Euclidean distances with a Pearson correlation coefficient of 0.904, which justifies our substitution.

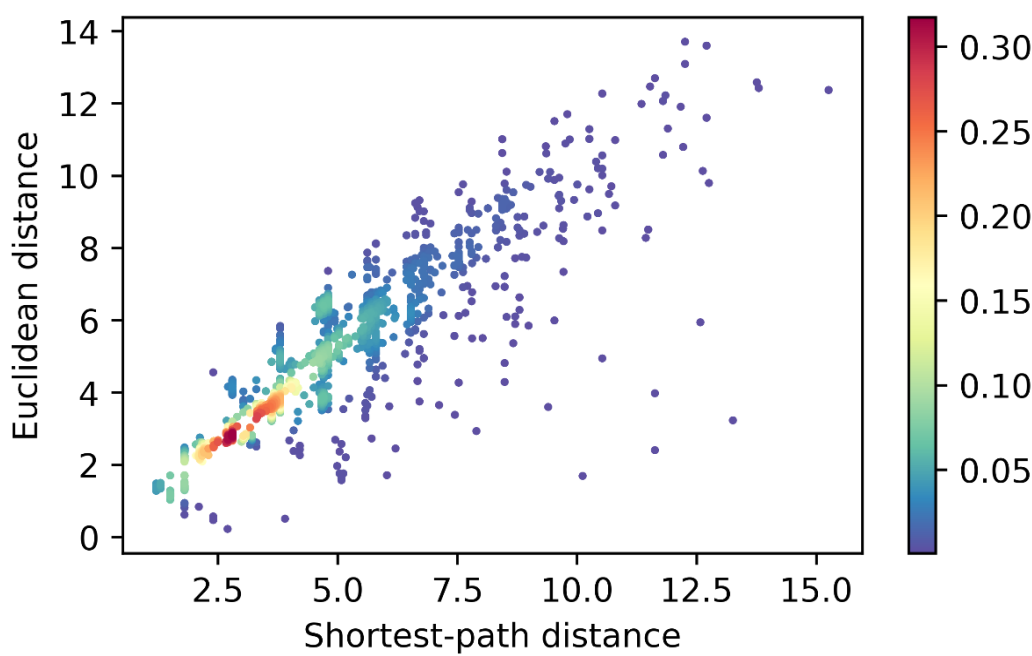

**Figure S1** | Euclidean and mapping distances between pharmacophore features.

**Calculation of match score**

A molecule may have many pharmacophore features and a subset of them can be used to construct a pharmacophore model. We can convert all the pharmacophore features of a molecule into a fully connected graph $G_f = \{V_f, E_f\}$ with anode set $V_f = \{v_{f_1}, v_{f_2}, ..., v_{f_n}\}$ and an edge set $E_m = \{e_{f_{1,2}}, e_{f_{1,3}}, ..., e_{f_{1,n}}, e_{f_{2,3}}, e_{f_{2,4}}, ..., e_{f_{n-1,n}}\}$. And a pharmacophore model be turned into a graph $G_p$ in the same way. Then the problem of calculating the matching degree between a given molecule and a given pharmacophore can be seen as finding the best match of a small graph in a large graph. Since molecules that we deal with usually contain a small number of heavy atoms, we simply use brute force to calculate the match score. The calculation steps are as follows:

**Input**: G: The pharmacophore graph to be matched; SMILES: SMILES to be verified

**Output**: The matching score for Pharmacophore and SMILES

**MATCH SCORE** (G, SMILES)

| | |
|---|---|
| 1 | V_r = G.nodes() |
| 2 | E_r=G.edges() |
| 3 | Extract chemical features from SMILES using RDKit |
| 4 | Transform SMILES into a graph G_Q(V_q, E_q) with chemical features as nodes |
| 5 | type_list = [ ] |
| 6 | score_list = [ ] |
| 7 | **for** i = 1 **to** length(V_r): |
| 8 |     type = [ ] |
| 9 |     **for** j=1 **to** length(V_r): |
| 10 |         **if** ref_type[i] == V_r[j]: |
| 11 |             type.append(V_r[j]) |
| 12 |     type_list.append(type) |
| 13 | **for** k = 1 **to** (length(type_list[0])* (length(type_list[1])*…* (length(type_list[-1]): |
| 14 |     dist_true=0 |
| 15 |     Extract one node and corresponding edge from different type_list at a time to get a candidate subgraph G_k(V_k, E_k) |
| 16 |     **for** l = 1 **to** length(V_r): |
| 17 |         **for** m = 1 **to** length( V_r): |
| 18 |             **if** m≠l **and** \|E_k(l, m)-E_r(l, m)\|<1.2: |
| 19 |                 dist_true = dist_true+1 |
| 20 |     score=dist_true/length(E_k) |
| 21 |     score_list.append(score) |
| 22 | **return** max(score_list) |

To make the calculation process understandable, we give some examples about the calculation of the match score. The pharmacophore in **Figure S2a** contains four pharmacophore features, where green indicates the hydrophobic center, red indicates the aromatic ring, and blue indicates the hydrogen-bonded acceptors. **Figure S2b**, **Figure S2c**, and **Figure S2d** give three molecules generated by PGMG with match scores of 1.0, 0.5, and 0.5, respectively. The molecule in **Figure S2b** is an example of a perfect match, while the molecules in **Figure S2c** and **Figure S2d** have their problems. In **Figure S2c**, the hydrogen-bonded acceptor formed by the carbonyl group is too far from the other pharmacophores. In **Figure S2d**, the molecule lacks a hydrogen-bonded acceptor and therefore can't match the lower right hydrogen-bonded acceptor in **Figure S2a**.

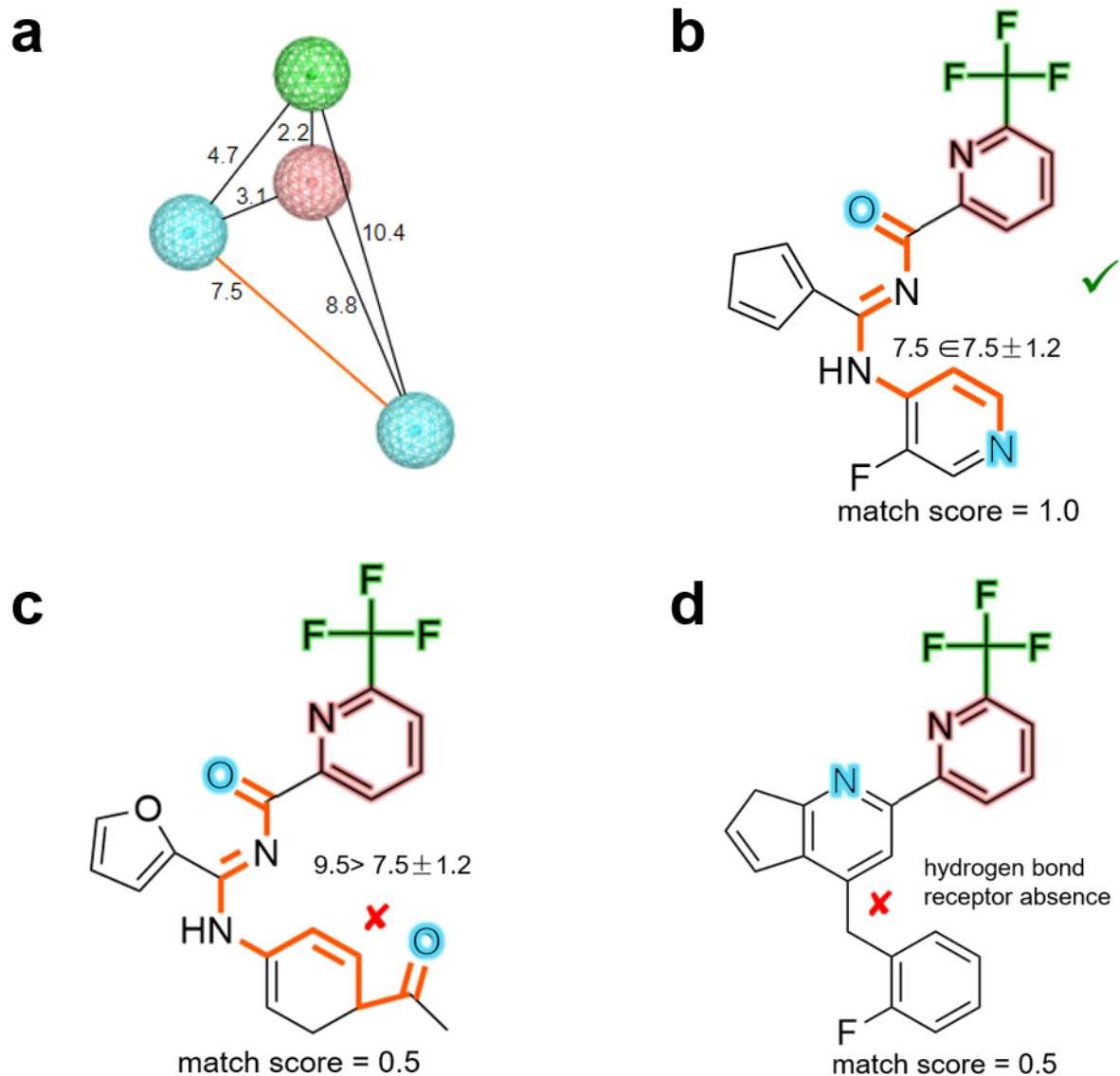

**Figure S2 | Illustration of the matching score calculation process.** (a) A pharmacophore model. (b) A molecule satisfying given pharmacophores. (c), (d) molecules unsatisfying given pharmacophores.

**Ablation study**

In the ablation study, we remove features of PGMG and see how that affects performance. All models are trained using the ZINC[2] dataset with the same parameters. Validity, uniqueness, and novelty are evaluated by generating 10 molecules for 1 pharmacophore extracted from each molecule in the test dataset. The match score is evaluated by generating 512×512 molecules for 512 SMILES randomly sampled from the test dataset. The result of our ablation study can be found in **Table S3**.

We find when using canonical SMILES to train PGMG (*canonical_SMILES*), the uniqueness increases from 0.976 to 0.991, but the match score decreases from 0.914 to 0.936. A similar result can be found when we change the Gaussian distribution of the latent variable *z* to a Dirac delta distribution, denoted as PGMG (*remove_z*). *remove_z* exhibits a huge decrease on the uniqueness (from 0.976 to 0.806) and a certain degree of increase on the match score (from 0.936 to 0.969). If we use random sampling during generation, we can make the uniqueness increase, but it cannot make up for the drop of both the validity and the match score. As we see here, there seems to be a trade-off between the uniqueness and match score.

We also test PGMG's performance when replacing the distance between chemical features with a constant number PGMG (*remove_dis*). The results show a large decrease in both uniqueness (from 0.976 to 0.823) and match score (from 0.936 to 0.596) as expected, which shows that PGMG makes a good use of the spatial information of pharmacophores.

Table S3. Results of the ablation study.

|  | Validity | Uniqueness | Novelty | Ratio of available molecules | Match score |
|---|---|---|---|---|---|
| PGMG | 0.972 | 0.976 | **0.998** | **94.8%** | 93.6% |
| PGMG(canonical_SMILES) | 0.976 | **0.991** | 0.997 | 96.4% | 91.4% |
| PGMG (remove_dis) | 0.990 | 0.823 | 0.996 | 81.2% | 59.6% |
| PGMG (remove_z) | **0.991** | 0.806 | 0.996 | 79.6% | **96.9%** |
| PGMG(remove_z, random sampling) | 0.917 | 0.999 | 0.999 | 91.5% | 88.8% |

**mapping**

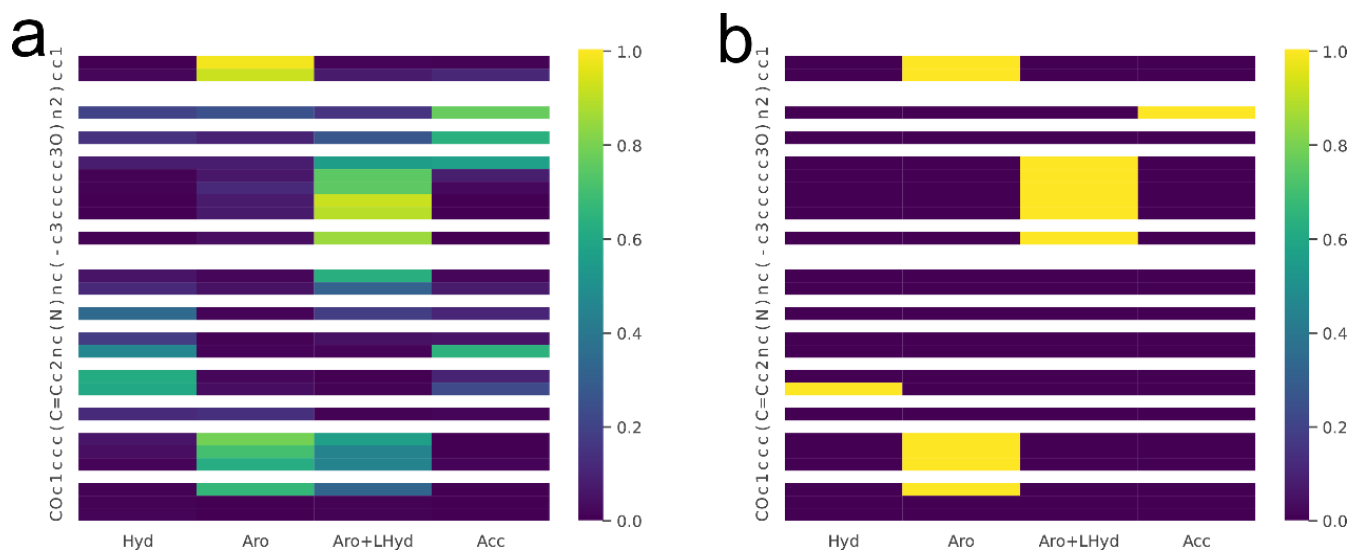

**Figure S3 | An illustration of masked mapping scores and labels. Tokens that are not heavy atoms are masked and denoted in white.** (a) masked mapping scores. (b) masked mapping labels.

**Demonstration of generated molecules**

**Table S4** shows the average affinity calculated using Autodock vina[3] of the top 1000 generated molecules and the affinity of the activity molecules with the specific targets obtained from the CHEMBL database. The average affinity of the top 1000 molecules with VEGFR2 and TGFβ 1 are not as good as the affinity of the reference ligand, while there are 70, 96 molecules with better affinity than the reference ligands. We calculate the RMSD of the top conformations of the reference ligand and the bioactive conformation acquired from

PDB[4]. And the RMSD shows that the top conformations of the reference ligand are relatively close to the active conformations acquired from PDB (**Figure S4**), which indicates the reliability of our docking results. The IC50 values of the reference ligands for specific targets are also listed in the table to facilitate a comparison of the molecule's inhibition of the target. It is also worth mentioning that the IC50 values of the four reference ligands corresponding to the proteins are all at the nanomolar level, and the same level of affinity of the generated molecules implies that the molecules generated by PGMG are also likely to have good biological activity.

Table S4. molecular docking results of different receptors generated by PGMG.

| target | PDB ID | average affinity of top 1000 molecules (kcal/mol) | average affinity of bioactivity molecules (kcal/mol) | affinity of reference ligand (kcal/mol) | RMSD (Å) | IC50 of reference ligand (nM) |
| --- | --- | --- | --- | --- | --- | --- |
| VEGFR2 | 1YWN | -10.0 | -8.0 | -10.8 | 0.59 | 3 |
| CDK6 | 2EUF | -11.1 | -9.6 | -10.5 | 1.36 | 15 |
| TGFβ 1 | 6B8Y | -11.0 | -9.2 | -11.5 | 0.45 | 0.56 |
| BRD4 | 3MXF | -8.8 | -7.0 | -8.0 | 0.80 | 49 |

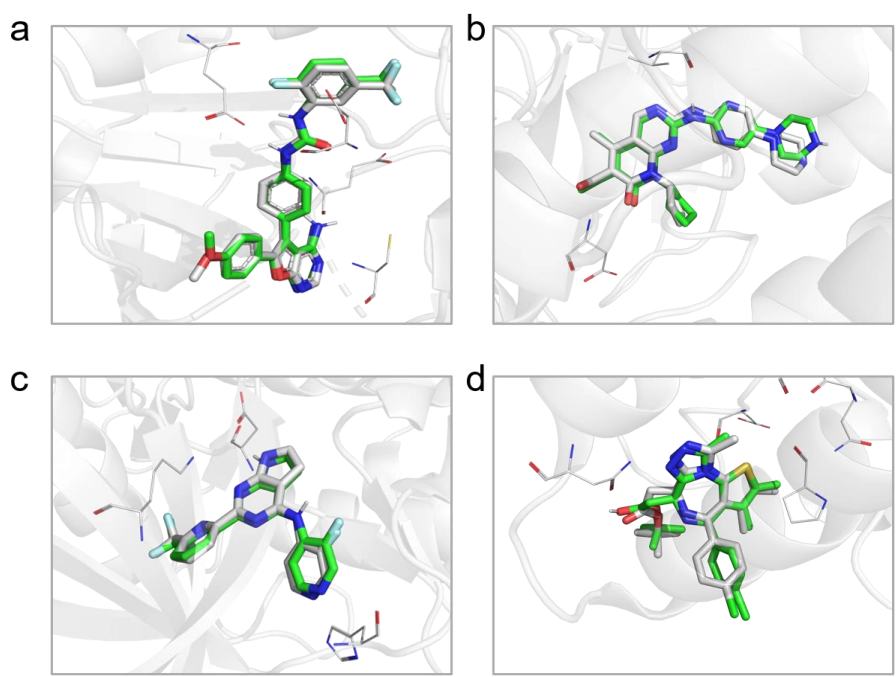

**Figure S4 | A display of the four reference ligand conformations in the PDB, and the top affinity conformation using vina docking in the protein pocket.** (a) the conformation in gray is the reference ligand of VEGFR2 (PDB:

1YWN), the conformation in green is acquired by autodock vina. (b) the conformation in gray is the reference ligand of CDK6 (PDB: 2EUF), the conformation in green is acquired by autodock vina. (c) the conformation in gray is the reference ligand of TGFβ 1 (PDB: 6B8Y), the conformation in green is acquired by autodock vina. (d) the conformation in gray is the reference ligand of BRD4 (PDB: 3MXF), the conformation in green is acquired by autodock vina.

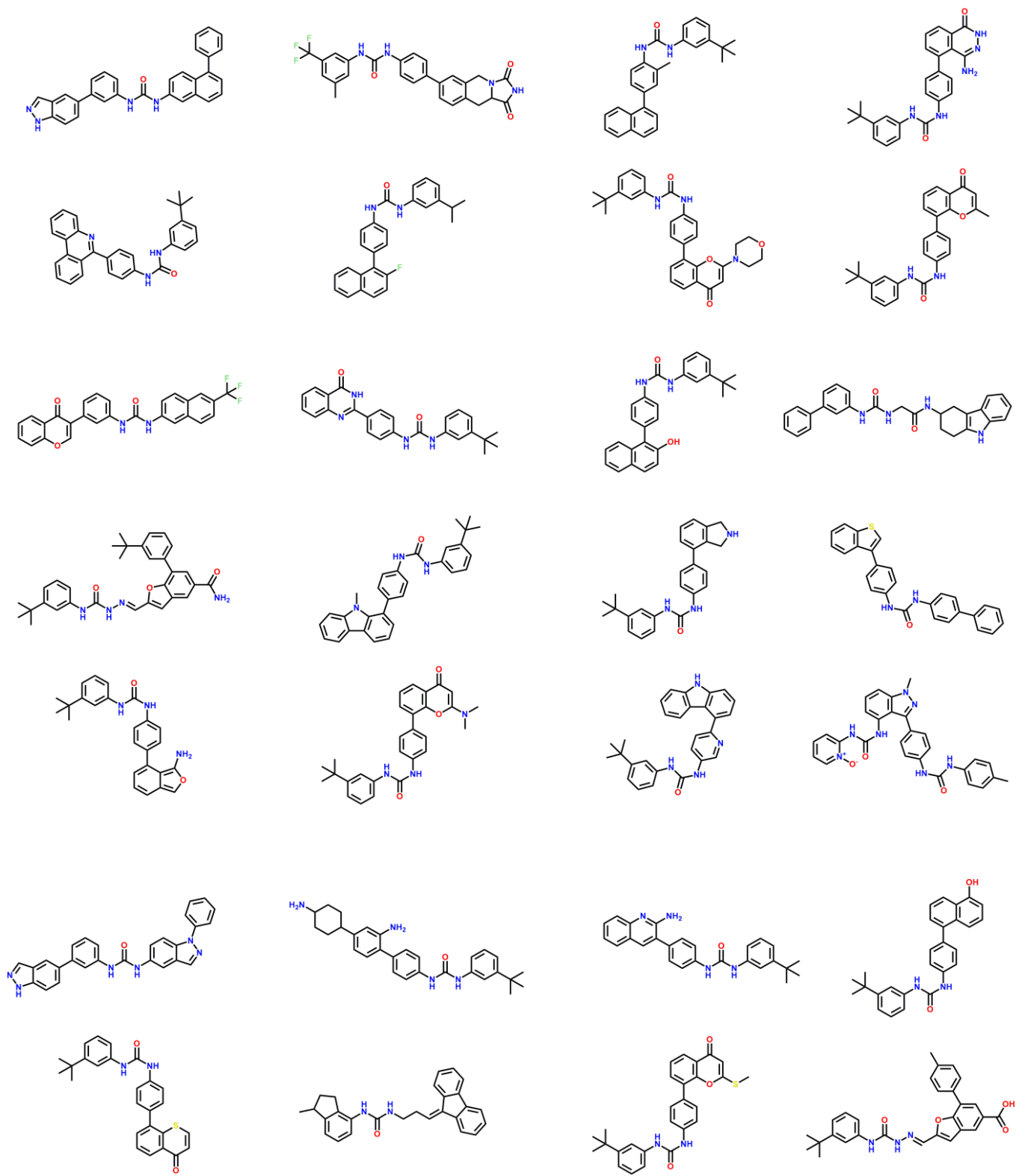

**Figure S5 | A display of 1YWN (PDB ID) generated novel molecules by PGMG.** The molecules are chosen from the generated molecules with top binding affinity.

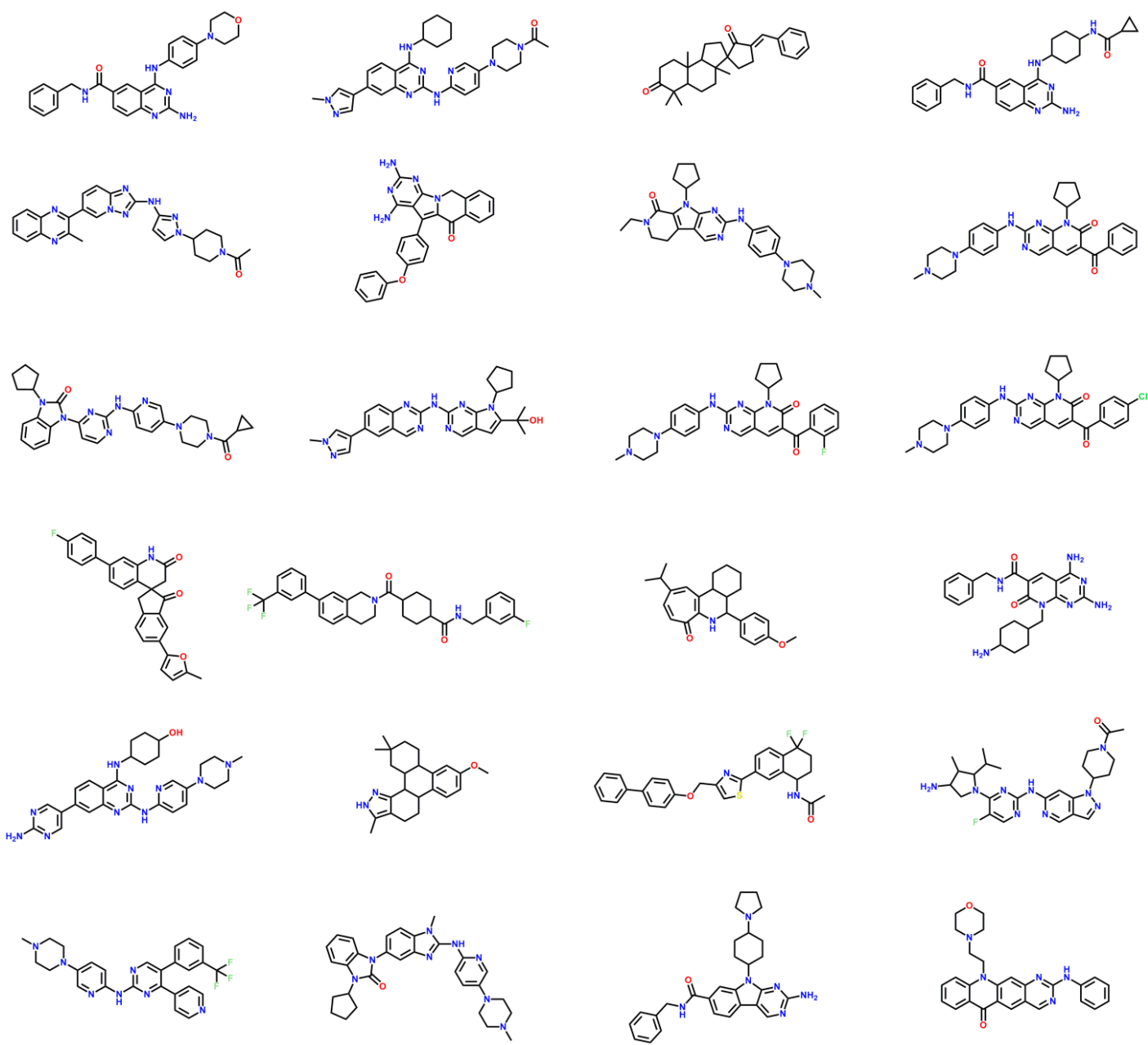

**Figure S6 | A display of 2EUF (PDB ID) generated novel molecules by PGMG.** The molecules are chosen from the generated molecules with top binding affinity.

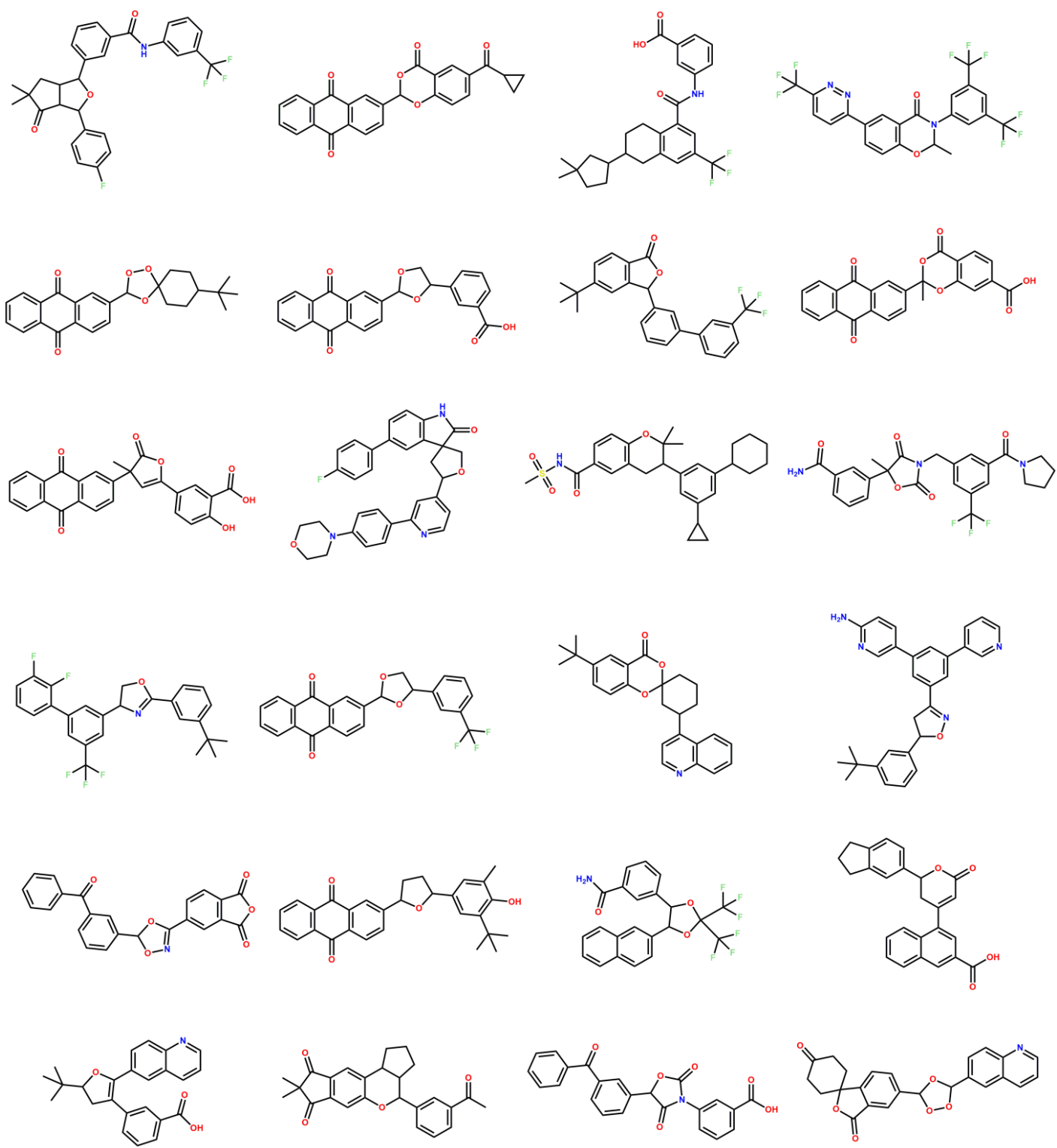

**Figure S7 | A display of 6B8Y ([PDB ID](#)) generated novel molecules by PGMG.** The molecules are chosen from the generated molecules with top binding affinity.

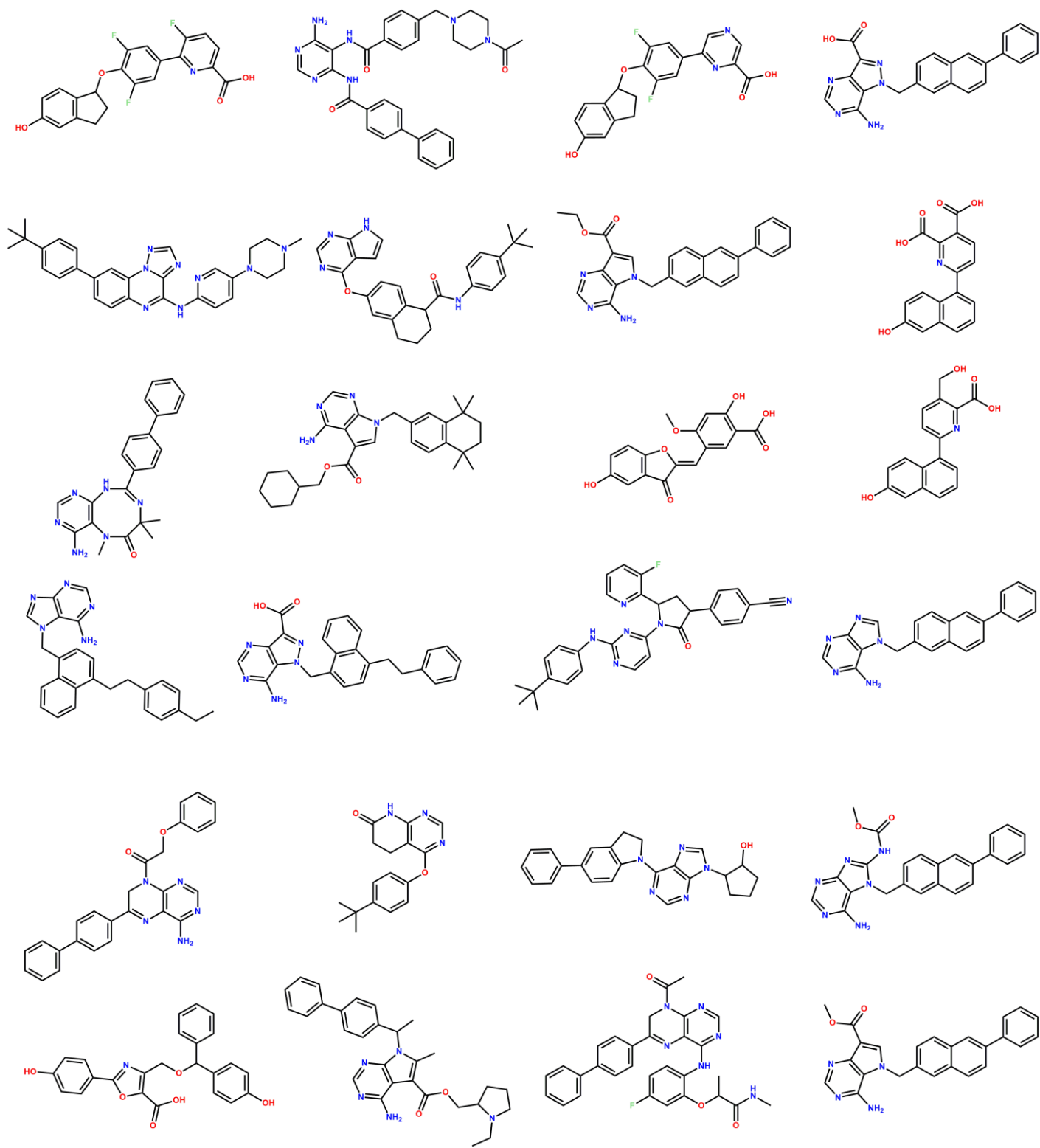

**Figure S8 | A display of 3MXF (PDB ID) generated novel molecules by PGMG.** The molecules are chosen from the generated molecules with top binding affinity.

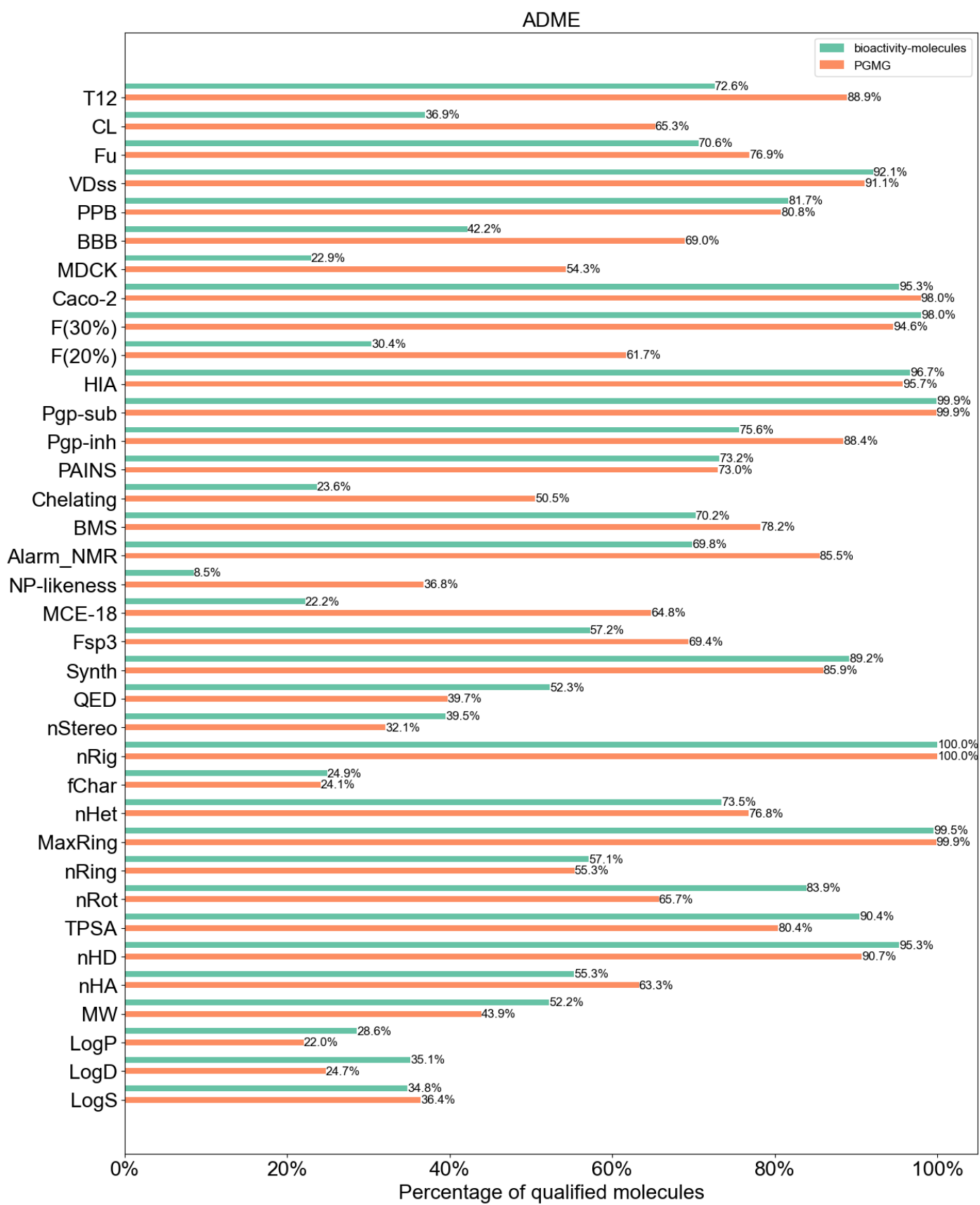

**Figure S9 | Percentage of qualified molecules of generated molecules and the known active molecules in terms of absorption, distribution, metabolism, excretion (ADME) properties.**

To assess the drug-like properties of generated molecules comprehensively, we use ADMETlab2.0[5] to calculate the absorption, distribution, metabolism excretion and toxicity properties of 21,618 active molecules from the CHEMBL database and 38,193 molecules generated according to four protein structures (VEGFR2,

CDK6, TGFβ 1, BRD4). We leave out indicators in the ADMETlab that are not given explicit thresholds and obtain 36 ADME indicators and 30 toxicity indicators.

**Figure S9** shows the percentage of molecules fulfilled the ADME properties. Overall, the molecules generated by PGMG behave comparably to the active molecules. The acquired bioactive molecules perform better in quantitative estimate of drug-likeness (QED), number of rotatable bonds (nRot) while the generated molecules outperform the active molecules by more than 10% among 9 indicators, including the half-life of a drug (T12), the clearance of a drug (CL), blood–brain barrier penetration (BBB), the human oral bioavailability 20% (F(20%)), the probability of being the inhibitor of P-glycoprotein (Pgp-inhibitor), thiol reactive compounds (Alarm-NMR), the natural product-likeness score (NP-likeness), medicinal chemistry evolution 2018 (MCE-18), and sp3 hybridized carbons/total (Fsp3).

**Figure S10** shows the percentage of qualified molecules that meet the toxicological criteria for PGMG generated molecules and the active molecules. The active molecules have advantages in the three toxicological criteria, NR-AhR, NR-PPAR-gamma, and SureChEMBL Rule (SureChEMBL). However, the generated molecules are significantly superior in 7 indicators including hERG Blockers (hERG), drug-induced liver injury (DILI), NR-AR-LBD, NR-Aromatase, skin sensitization rule (Skin_Sensitization), and acute toxicity rule (A_A_Toxicity). The above results demonstrate that PGMG has the ability to generate drug-like molecules.

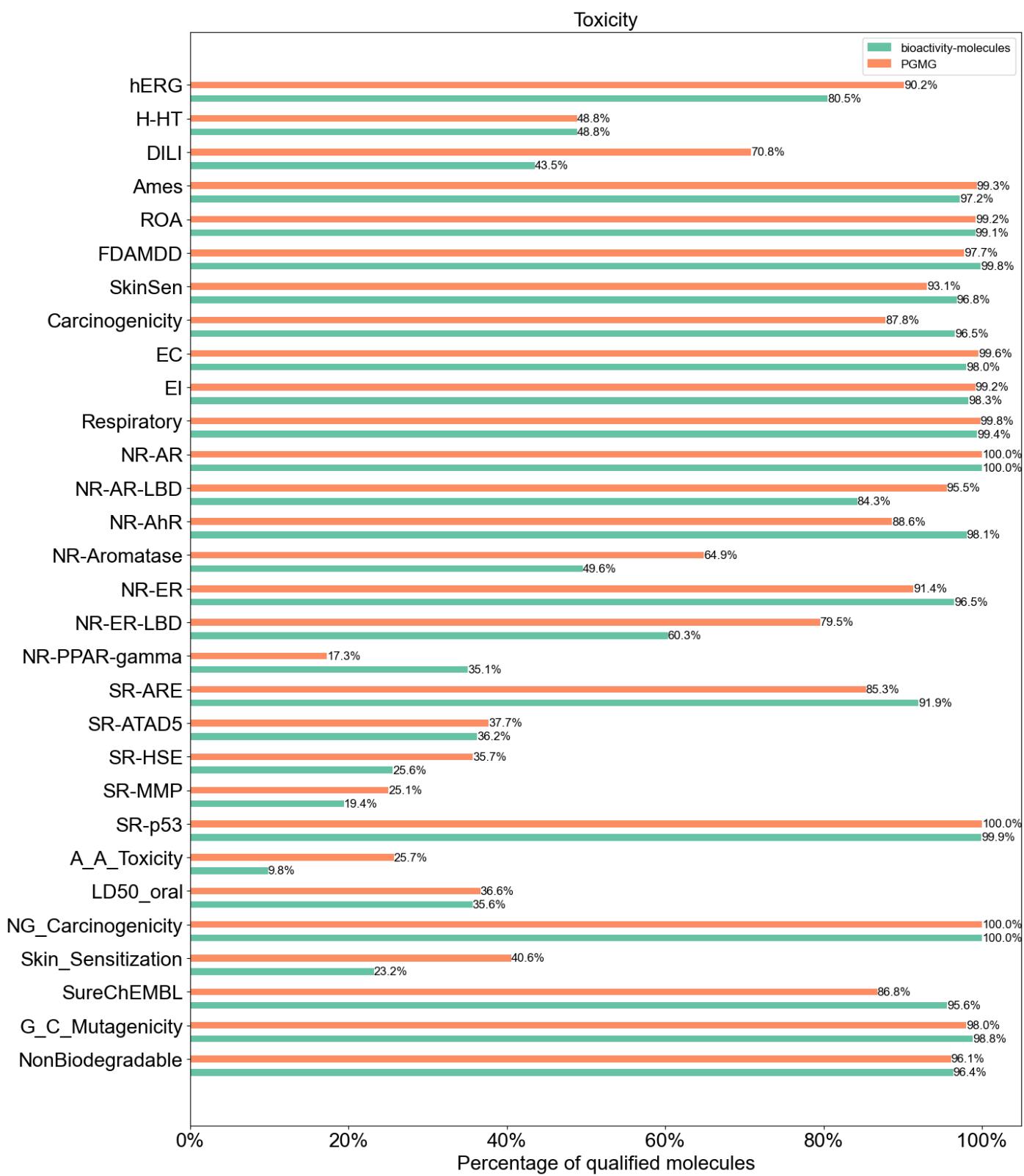

**Figure S10 | Percentage of qualified molecules of PGMG-generated molecules and the known active molecules in terms of toxicological indicators.**

reference

1. Landrum G. RDKit: Open-source cheminformatics. http://www.rdkit.org.
2. Sterling T, Irwin JJ. ZINC 15–ligand discovery for everyone. *Journal of chemical information and modeling* **55**, 2324-2337 (2015).
3. Trott O, Olson AJ. AutoDock Vina: improving the speed and accuracy of docking with a new scoring

function, efficient optimization, and multithreading. *Journal of computational chemistry* **31**, 455-461 (2010).
4. Burley SK, Berman HM, Kleywegt GJ, Markley JL, Nakamura H, Velankar S. Protein Data Bank (PDB): the single global macromolecular structure archive. *Protein Crystallography* **1607**, 627-641 (2017).
5. Xiong G*, et al.* ADMETlab 2.0: an integrated online platform for accurate and comprehensive predictions of ADMET properties. *Nucleic Acids Research* **49**, W5-W14 (2021).